\RequirePackage[T1]{fontenc}
\documentclass[12pt]{article}
\usepackage[height=8.85in,width=6.45in]{geometry}

\usepackage[utf8]{inputenc}

\usepackage{amsmath}
\usepackage{amssymb}
\usepackage{mathtools}
\usepackage{amsthm}

\usepackage{times}
\usepackage[scaled]{couriers}
\usepackage{mathrsfs}

\usepackage{graphicx}
\usepackage{subcaption}

\usepackage{tikz}
\usepackage{tikz-cd}
\usetikzlibrary{calc}
\usetikzlibrary{matrix,positioning}

\usepackage[svgnames]{xcolor}
\usepackage[colorlinks,linktocpage=true,citecolor=DarkGreen,linkcolor=FireBrick]{hyperref}
\usepackage{cite}

\definecolor{CUDred    }{RGB}{255, 75,  0}\def\colorA{CUDred    }
\definecolor{CUDorange }{RGB}{246,170,  0}\def\colorB{CUDorange }
\definecolor{CUDgreen  }{RGB}{  3,175,122}\def\colorC{CUDgreen  }
\definecolor{CUDskyblue}{RGB}{ 77,196,255}\def\colorD{CUDskyblue}
\definecolor{CUDblue   }{RGB}{  0, 90,255}\def\colorE{CUDblue   }
\definecolor{CUDpurple }{RGB}{153,  0,153}\def\colorF{CUDpurple }
\definecolor{CUDpink   }{RGB}{255,128,130}\def\colorG{CUDpink   }
\definecolor{CUDgray   }{RGB}{132,145,158}\def\colorH{CUDgray   }
\definecolor{CUDbrown  }{RGB}{128, 64,  0}\def\colorI{CUDbrown  }
\definecolor{CUDyellow }{RGB}{255,241,  0}

\usepackage{bm}
\usepackage{slashed}
\usepackage{braket}

\numberwithin{equation}{section}

\theoremstyle{plain}

\theoremstyle{definition}

\numberwithin{thm}{section}

\usepackage{spectralsequences}

\def\d{{\rm d}}
\def\i{{\mathsf i}}

\def\Ker{\mathop{\mathrm{Ker}}}

\def\Hom{\mathop{\mathrm{Hom}}}

\def\cA{{\cal A}}

\def\cD{{\cal D}}

\def\cN{{\cal N}}
\def\cO{{\cal O}}

\def\bP{{\mathbb P}}
\def\bQ{{\mathbb Q}}
\def\bR{{\mathbb R}}

\def\bZ{{\mathbb Z}}

\def\O{\mathrm{O}}
\def\SO{\mathrm{SO}}

\def\USp{\mathrm{USp}}
\def\Spin{\mathrm{Spin}}
\def\Pin{\mathrm{Pin}}

\usepackage{amsmath}
\usepackage{mathtools}

\DeclareMathOperator{\Sq}{Sq}
\def\pt{\mathrm{pt}}
\def\DPin{\mathrm{DPin}}

\begin{document}

\begin{titlepage}

\begin{flushright}
YITP-26-117\\
KEK-TH-2867\\
IPMU26-0031\\
TU-1319
\end{flushright}

\vskip 3cm

\begin{center}

{\large \bfseries Anomalies and discrete torsion in Type II and Type I worldsheet theories}

\vskip 1cm
 Masashi~Kawahira$^1$, 
 Shotaro~Kawanago$^2$,
 Shota~Saito$^3$,
 and Hiroki~Wada$^4$
\vskip 1cm

\begin{tabular}{ll}
$^1$ & Yukawa Institute for Theoretical Physics, Kyoto University, Kyoto 606-8502, Japan\\
$^2$ & Graduate Institute for Advanced Studies, SOKENDAI, 1-1 Oho, Tsukuba, \\
& Ibaraki 305-0801, Japan\\
$^3$ & Kavli Institute for the Physics and Mathematics of the Universe (WPI),
University of Tokyo, \\
& Kashiwa, Chiba 277-8583, Japan\\
$^4$ & Department of Physics, Tohoku University, Sendai 980-8578, Japan\\
\end{tabular}

\vskip 1cm

\end{center}

\noindent
We provide a comprehensive study of anomalies and discrete torsion in the worldsheet theories of Type~II and Type~I string theories on a general ten-dimensional target space.
In the modern understanding, the anomaly of a worldsheet theory is characterized by an invertible field theory in three dimensions, and invertible field theories are classified by the Anderson dual of a suitable bordism theory.
To study the anomaly cancellation condition, we employ the explicit model of the Anderson dual constructed by Yamashita and Yonekura.
We find that, for both Type~II and Type~I string theories, the anomaly of the worldsheet theory is cancelled if and only if the target space is orientable and admits a spin structure.
We also determine the discrete torsion and clarify its physical meaning.
It includes the distinctions between Type~IIA and Type~IIB string theories and between ordinary Type~I and Sugimoto string theories.

\end{titlepage}

\setcounter{tocdepth}{3}

\tableofcontents

\newpage

\section{Introduction and summary}

In the worldsheet formulation of string theory, anomalies give rise to important consistency conditions.
The most elementary example is the critical dimension.
The cancellation of the conformal anomaly fixes the central charge of the matter sector, and hence the dimension of the target space.
More generally, a worldsheet theory in the Neveu--Schwarz--Ramond formalism~\cite{Neveu:1971rx,Ramond:1971gb} may suffer from an obstruction to the Gliozzi--Scherk--Olive (GSO) projection~\cite{Gliozzi:1976jf,Gliozzi:1976qd} as well as from a sigma-model anomaly~\cite{Moore:1984dc,Moore:1984ws}.
The cancellation of these anomalies imposes nontrivial restrictions on the topology of the target space.
The first purpose of this paper is to study the relation between such anomalies and the topology of the target space for Type~II and Type~I string theories, which are based on $\cN=(1,1)$ worldsheet theories.

From the modern point of view, the anomaly of a $d$-dimensional theory is characterized by the partition function of a $(d+1)$-dimensional invertible field theory~\cite{Freed:2004yc} through the anomaly inflow mechanism~\cite{Callan:1984sa}.\footnote{The application of the anomaly inflow mechanism to the conformal anomaly involves difficulties. See Ref.~\cite{Aminov:2026zrv} for related discussions.}
For fermions, the invertible field theory is constructed by realizing the $d$-dimensional fermions as boundary modes of massive fermions in one higher dimension, and its partition function is expressed in terms of the $\eta$-invariant of the corresponding Dirac operator~\cite{Witten:2015aba,Witten:2019bou}.
It is moreover believed that invertible field theories are classified by the Anderson dual of a suitable bordism theory~\cite{Kapustin:2014tfa,Kapustin:2014dxa,Freed:2016rqq,Yonekura:2018ufj}.
For a two-dimensional worldsheet theory with target space $X$, the anomaly is accordingly classified by the group $(I_{\bZ}\Omega^{H})^{4}(X)$, where $H$ is the tangential structure of the worldsheet.
This tangential structure encodes how the GSO projection or the orientifold projection~\cite{Sagnotti:1987tw,Dai:1989ua} is implemented.
In the cases considered in this paper, the tangential structures are given by $H=\Spin\times\bZ_{2}$ for Type~II string theory and by $H=\DPin$ for Type~I string theory.

Once the anomaly is cancelled, the worldsheet theory is still not completely specified.
One can add a topological term, and the choice of such a term is part of the data defining the theory.
In the context of worldsheet theories, this choice is referred to as the discrete torsion~\cite{Vafa:1986wx}.
Any two choices differ by an element of $\Hom(\Omega^{H}_{2}(X),\bR/\bZ)$, so that the possible choices form a torsor over this group.
Physically, part of the discrete torsion corresponds to a choice of topological data of the target space, such as an orientation and a spin structure.
Furthermore, the other part of the discrete torsion distinguishes different string theories.
For instance, Type~IIA and Type~IIB string theories differ by a choice of the discrete torsion.
The flat part of the $B$-field is also included in the discrete torsion.
The second purpose of this paper is to clarify the physical meaning of the discrete torsion for Type~II and Type~I string theories.

The relation between worldsheet anomalies and the topology of the target space has been studied extensively in various contexts.
In Ref.~\cite{Freed:1999vc}, the anomaly of the Type~II worldsheet theory was analyzed in terms of the Pfaffian line bundle.
The interplay between the anomaly and the $B$-field was discussed in Ref.~\cite{Witten:1999eg}.
More recently, several analyses based on bordism theory have been carried out.
For a flat target space, Type~0~\cite{Dixon:1986iz}, Type~0 orientifolds~\cite{Bianchi:1990yu,Sagnotti:1996qj,Bergman:1997rf,Bergman:1999km,Blumenhagen:1999ad}, Type~II, and Type~I string theories were studied in Refs.~\cite{Kaidi:2019pzj,Kaidi:2019tyf,Witten:2023snr}.
For a general target space, the anomaly cancellation condition and the physical meaning of the discrete torsion of Type~II string theory were determined in Ref.~\cite{Delgado:2026qvy}.
An analogous analysis was performed in Ref.~\cite{Yonekura:2022reu} for supersymmetric quantum mechanics and for a class of heterotic string theories~\cite{Gross:1984dd,Gross:1985fr,Gross:1985rr} by employing the explicit model of the Anderson dual constructed by Yamashita and Yonekura~\cite{Yamashita:2021cao}.
In this paper, we reproduce the known results for Type~II string theory by a route different from that of Ref.~\cite{Delgado:2026qvy}, and carry out the corresponding analysis for Type~I string theory on a general target space.

Our analysis is based on the explicit model of the Anderson dual of bordism theory constructed by Yamashita and Yonekura~\cite{Yamashita:2021cao}, in which a class is represented by a pair $(h,\omega)$ consisting of an $\bR/\bZ$-valued bordism invariant and a closed four-form.
We follow the strategy developed in Ref.~\cite{Yonekura:2022reu} and apply it to Type~II and Type~I string theories.
The group $(I_{\bZ}\Omega^{H})^{4}(X)$ fits into a short exact sequence analogous to the universal coefficient theorem: the perturbative anomaly is captured by $\Hom(\Omega^{H}_{4}(X),\bZ)$, while the purely global anomaly is encoded in $\mathrm{Ext}(\Omega^{H}_{3}(X),\bZ)$.
In the cases of our interest, the anomaly polynomial vanishes, so that the anomaly is determined by the values of $h$ on the torsion subgroup of $\Omega^{H}_{3}(X)$.
Note that the worldsheet fermions couple to the tangent bundle of the target space through the sigma model-field, and that the tangent bundle is obtained as the pullback of the universal bundle along the classifying map $f:X\to B\O(10)$.
It follows that the anomaly lies in the image of $f^{\ast}:(I_{\bZ}\Omega^{H})^{4}(B\O(10))\to(I_{\bZ}\Omega^{H})^{4}(X)$.
Therefore, it is sufficient in practice to compute $\Omega^{H}_{3}(B\O(10))$ rather than $\Omega^{H}_{3}(X)$ for each target space.
This is where our derivation differs from that of Ref.~\cite{Delgado:2026qvy}, in which the relevant bordism groups for a general target space are computed directly by means of the Smith isomorphism.
The two bordism groups we need are $\Omega^{\Spin\times\bZ_{2}}_{3}(B\O(10))$ and $\Omega^{\DPin}_{3}(B\O(10))$, which we compute by the Adams spectral sequence in App.~\ref{app:bordism}.
For the discrete torsion, on the other hand, no such reduction is available, and we work directly with $\Omega^{H}_{2}(X)$ for a general target space $X$.

For Type~II string theory, we find that the anomaly of the worldsheet theory is cancelled if and only if the target space is orientable and admits a spin structure.
This result is, of course, the same as that obtained in Ref.~\cite{Delgado:2026qvy}.
The discrete torsion is then parametrized by $\Hom(\Omega^{\Spin\times\bZ_{2}}_{2}(X),\bR/\bZ)$, which is built out of the following pieces.
There are two $\bZ_{2}$ summands generated by the Arf invariants of the spin structures in the left- and right-moving sectors.
One distinguishes Type~IIA from Type~IIB string theory, while the other determines the orientation of the target space.
In addition, there are two copies of $H^{1}(X;\bZ_{2})$, which serve as background fields for the quantum symmetries arising from gauging the left- and right-moving fermion number symmetries.
Their diagonal combination is the spacetime fermion number symmetry, and turning on its background field shifts the spin structure of the target space.
The last piece is $H^{2}(X;\bR/\bZ)$, which is the flat part of the $B$-field.

For Type~I string theory, the anomaly cancellation condition turns out to be the same as that of Type~II string theory.
Namely, the target space is again required to be orientable and to admit a spin structure.
The discrete torsion, on the other hand, differs from that of Type~II string theory in three respects.
First, the $\bZ_{2}$ that distinguishes Type~IIA from Type~IIB string theory is replaced by the topological term $w_{1}(T\Sigma)^{2}$ built from the worldsheet $\Sigma$, which distinguishes ordinary Type~I string theory from Sugimoto string theory~\cite{Sugimoto:1999tx}.
The $\bZ_{2}$ summand specifying the orientation of the target space, however, is still present.
Second, the two copies of $H^{1}(X;\bZ_{2})$ are reduced to a single one, since the orientifold projection exchanges the quantum symmetries for the left- and right-moving sectors and only their diagonal combination survives.
Third, the flat part of the $B$-field is reduced from $H^{2}(X;\bR/\bZ)$ to $H^{2}(X;\bZ_{2})$, in accordance with the fact that the orientifold projection removes the $B$-field from the massless spectrum while leaving a $\bZ_{2}$-valued degree of freedom~\cite{Bianchi:1991eu,Sen:1997pm,Witten:1997bs}.

The organization of this paper is as follows.
In Sec.~\ref{sec:Generality}, we review the formulation and the classification of fermion anomalies in the model of Ref.~\cite{Yamashita:2021cao}, and formulate the anomaly cancellation condition for worldsheet theories.
In Sec.~\ref{sec:Type_II}, we analyze the anomaly and the discrete torsion of Type~II string theory.
In Sec.~\ref{sec:Type_I}, we first review the $\DPin$ structure and then carry out the corresponding analysis for Type~I string theory.
We conclude with discussions in Sec.~\ref{sec:Discussion}.
The bordism groups used in the main text are computed in App.~\ref{app:bordism}.

\section{Generalities}
\label{sec:Generality}

In this section, we review the modern perspective on sigma-model anomalies and develop the framework leading to \eqref{eq:global_anomaly_cancellation}.
In Sec.~\ref{subsec:Yamashita-Yonekura}, we present the general formulation of anomalies in terms of the Anderson dual of bordism groups, together with its differential extension.
In Sec.~\ref{subsec:anomaly_cancellation}, we apply this formulation to sigma-model anomalies in worldsheet theories.

\subsection{Anomalies as invertible QFT}
\label{subsec:Yamashita-Yonekura}

From the modern perspective, the anomaly of a $d$-dimensional theory is characterized by the partition function of a $(d+1)$-dimensional invertible field theory~\cite{Freed:2004yc} based on the anomaly inflow mechanism~\cite{Callan:1984sa}.
In addition, it is conjectured that such invertible field theories are classified by the Anderson dual of bordism groups~\cite{Kapustin:2014tfa,Kapustin:2014dxa,Freed:2016rqq,Yonekura:2018ufj}.
Yamashita and Yonekura constructed an explicit model of the Anderson dual group and its differential extension~\cite{Yamashita:2021cao}, which supports this conjecture.
With these formulations, anomalies of fermions are better understood and can be described non-perturbatively in terms of the $\eta$-invariant and index densities~\cite{Witten:2015aba,Witten:2019bou}.

In the presence of an anomaly, the partition function is not a well-defined function, but rather a section of a line bundle.
Note that this situation is the same for sigma-model anomalies~\cite{Moore:1984dc,Moore:1984ws}.
To make the partition function well-defined, one introduces a bulk theory in one higher dimension and cancels the ambiguity via anomaly inflow.
Then, an anomaly is encoded in the dependence on the extension into the bulk.
In our case, let us consider two three-manifolds $W'$ and $W''$ with a common boundary $\Sigma$, which we regard as a worldsheet.
The anomaly is captured by the bulk partition function
\begin{align}\begin{aligned}
    Z(W)&=\exp\left(2\pi\i h(W)\right),&
    &h(W)\in\bR/\bZ,
\end{aligned}\end{align}
where $W$ is the closed three-manifold $W=W'\cup\overline{W''}$ obtained by gluing $W'$ and $\overline{W''}$ along $\Sigma$, where $\overline{W''}$ denotes the orientation reversal of $W''$.\footnote{For an unoriented tangential structure such as $\DPin$, the orientation reversal in this gluing is to be understood as the reversal of the $H$-structure induced by reversing the normal direction along the boundary. See Refs.~\cite{Freed:2016rqq,Yonekura:2018ufj} for details.}
For fermion anomalies, $h(W)$ is expressed in terms of the $\eta$-invariant~\cite{Dai:1994kq}.
In the applications below, we are concerned with Majorana--Weyl fermions, whose contribution to the $\eta$-invariant is half that of Dirac fermions.
We accordingly have $h(W)=-\eta(\cD_{W})/2$, where $\cD_{W}$ is the Dirac operator for the massive fermions on $W$.

The anomaly, viewed as an ambiguity in the partition function, is naturally understood in the differential model of the Anderson dual of bordism groups $(\widehat{I_\bZ\Omega^H})^4(X)$ introduced in~\cite{Yamashita:2021cao}.
Here, we have included the target space $X$ for later use.
Elements of the differential group are given by pairs $(h,\omega)$, where $\omega$ is a closed four-form on $X$ taking values in invariant polynomials of the Lie algebra of $H$, and $h$ is a map which assigns a value in $\bR/\bZ$ to a pair $(W,\phi)$ with $\phi:W\to X$.
Now, suppose that the three-manifold $W$ bounds a four-manifold $L$, to which all the background data are extended.
Then, we have a compatibility condition
\begin{align}
    h(\partial L)=\int_{L}\,\omega\quad \mod\bZ.
\end{align}
At the topological level, we should take into account the effect of local counterterms.
A local counterterm that does not contribute to the anomaly inflow is constructed from a manifestly gauge-invariant $3$-form $\alpha$.
The associated counterterm on the three-manifold $W$ is given by
\begin{align}
    h_{\alpha}(W):=\int_{W}\,\alpha.
\end{align}
For the purpose of classifying anomalies, the invertible field theory corresponding to the pair $(h_{\alpha},\d\alpha)$ must be regarded as trivial.
After this identification, we obtain the classification of the deformation class of invertible field theories $(I_\bZ\Omega^H)^4(X)=(\widehat{I_\bZ\Omega^H})^4(X)/\sim$.
We denote the equivalence class of $(h,\omega)$ by $[(h,\omega)]$.
In summary, $h$ corresponds to the partition function of the invertible field theory, while $\omega$ corresponds to an anomaly polynomial~\cite{Zumino:1983rz,Alvarez-Gaume:1983ihn,Alvarez-Gaume:1984zlq}.

A particularly useful feature of $(I_{\bZ}\Omega^{H})^{4}(X)$ is that it fits into the short exact sequence
\begin{align}
\label{eq:UCT}
0
\to
\mathrm{Ext}\left(\Omega^{H}_{3}(X),\bZ\right)
\xrightarrow{}
(I_{\bZ}\Omega^H)^{4}(X)
\xrightarrow{p}
\Hom\left(\Omega^{H}_{4}(X),\bZ\right)
\to
0.
\end{align}
This sequence includes the effects of both perturbative and global anomalies.
The map $p$ sends a class $[(h,\omega)]$ to the homomorphism $[L]\mapsto\int_{L}\omega$.
The image of a class under $p$ retains only the information about the anomaly polynomial $\omega$, which corresponds to the perturbative part.
On the other hand, an element of $\Ker p$ can always be represented by a pair of the form $(h,0)$ which describes a purely global anomaly, and the exactness of the sequence implies that it belongs to $\mathrm{Ext}\left(\Omega^{H}_{3}(X),\bZ\right)$.
We note that the sequence~\eqref{eq:UCT} does not split canonically, and consequently the decomposition of a general anomaly into perturbative and global parts is not canonical either.
For later use, it is convenient to state the condition under which an element of $\Ker p$ is trivial.
Since $\omega$ vanishes, $h$ is a bordism invariant and hence defines a homomorphism from $\Omega^{H}_{3}(X)$ to $\bR/\bZ$.
One can show that $[(h,0)]$ vanishes if and only if $h$ vanishes on the torsion subgroup of $\Omega^{H}_{3}(X)$.

Finally, the above construction of the partition function on $\Sigma$ presupposes the existence of a three-manifold $W'$ bounded by $\Sigma$.
When $\Sigma$ is not null-bordant as an element of $\Omega^{H}_{2}(X)$, this prescription is unavailable, and a phase must be assigned to each bordism class as part of the definition of the theory.
Any two such assignments differ by an element of $\Hom(\Omega^{H}_{2}(X),\bR/\bZ)$.
These phases are the generalized $\theta$-angles associated with the background fields, and we shall refer to them as discrete torsion~\cite{Vafa:1986wx} in the following sections.

\subsection{Anomalies in worldsheet theories}
\label{subsec:anomaly_cancellation}

Here, we review the anomaly cancellation condition for the worldsheet theory.
String theory is formulated using a sigma-model map from the worldsheet to the target space.
Therefore, we may encounter sigma-model anomalies~\cite{Moore:1984dc,Moore:1984ws}, and must verify that such anomalies vanish in order to treat the sigma-model map as a dynamical field.
A general framework based on the bordism point of view is discussed in Ref.~\cite{Yonekura:2022reu}, and in this paper we follow this approach.

First of all, let us construct a worldsheet theory which is an $\cN=(1,1)$ supersymmetric sigma model with a ten-dimensional target space $X$.
Note that we consider the chiral GSO projection and choose different spin structures for left- and right-moving fermions.
Therefore, instead of a spin structure, we consider an $H=\Spin\times\bZ_{2}$ structure for Type~II string theory and an $H=\DPin$ structure (see Sec.~\ref{subsec:DPin}) for Type~I string theory.
Now, the scalar field is described by a sigma-model field $\phi:\Sigma\to X$, while the fermions are sections of the pullback $\phi^{\ast}TX$ of the tangent bundle.
More precisely, the Dirac operator is constructed from the pullback of the Levi-Civita connection on $X$, together with the metric and the $H$-structure on $\Sigma$.
As reviewed in Sec.~\ref{subsec:Yamashita-Yonekura}, the anomaly in the worldsheet theory then specifies an element of $(I_{\bZ}\Omega^{H})^{4}(X)$.

The key point is that the anomalies actually realized in the worldsheet theory are captured by a subgroup of $(I_{\bZ}\Omega^{H})^{4}(X)$, for the following reason.
Since the tangent bundle of $X$ is classified by a map $f:X\to B\O(10)$, where $B\O(10)$ is the classifying space of $\O(10)$, the worldsheet fermions couple to the rank-10 real vector bundle associated with the composite map $f\circ\phi:\Sigma\to B\O(10)$.
This observation leads to the following simplification.
Instead of the original sigma model, we may consider a theory with an $\O(10)$ symmetry.
Its background data consist of the metric, the $H$-structure, and a connection on a rank-10 real vector bundle on $\Sigma$, and the fermions now couple directly to this connection rather than through the sigma-model field.
The anomaly of this theory is specified by an element $[(h_{0},\omega_{0})]\in(I_{\bZ}\Omega^{H})^{4}(B\O(10))$, and the anomaly of the original sigma model is obtained by the pullback $f^{\ast}[(h_{0},\omega_{0})]$.\footnote{The Yamashita--Yonekura model for $(I_{\bZ}\Omega^H)^n(X)$ is formulated when $X$ is a manifold. Hence, to study $(I_{\bZ}\Omega^H)^n(B\O(10))$, we may use finite-dimensional manifold approximations to $B\O(10)$.}
The anomaly cancellation condition can therefore be written as
\begin{align}\label{eq:anomaly_cancellation}
    f^{\ast}[(h_{0},\omega_{0})]=0\in(I_{\bZ}\Omega^{H})^{4}(X).
\end{align}
As we will see in Secs.~\ref{sec:Type_II} and~\ref{sec:Type_I}, the perturbative anomalies are cancelled in the cases considered in this paper; hence $\omega_{0}$ can be taken to be zero, in which case $h_{0}$ is regarded as an element of $\Hom(\Omega^{H}_{3}(B\O(10)),\bR/\bZ)$.
Recall that an element of $\Omega^{H}_{3}(B\O(10))$ is a bordism class of pairs $(W,g)$ consisting of a three-manifold $W$ with an $H$-structure and a classifying map $g:W\to B\O(10)$, and that $h_{0}$ assigns a number $h_{0}(W,g)$ to each such pair.
Given a bordism class of $\Omega^{H}_{3}(X)$ represented by a pair $(W,\phi)$ consisting of a three-manifold $W$ with an $H$-structure and a map $\phi:W\to X$, we take
\begin{align}
    f^{\ast}h_{0}(W,\phi)=h_{0}(W,f\circ\phi),
\end{align}
which yields an element of $\Hom(\Omega^{H}_{3}(X),\bR/\bZ)$.
We then have $f^{\ast}[(h_{0},0)]=[(f^{\ast}h_{0},0)]$, so that the anomaly cancellation condition reduces to
\begin{align}\label{eq:global_anomaly_cancellation}
    [(f^{\ast}h_{0},0)]=0\in(I_{\bZ}\Omega^{H})^{4}(X).
\end{align}
As explained in Sec.~\ref{subsec:Yamashita-Yonekura}, this condition is equivalent to the vanishing of $f^{\ast}h_{0}$ on the torsion subgroup of $\Omega^{H}_{3}(X)$.
\section{Type II string theory}
\label{sec:Type_II}

Based on the framework reviewed in Sec.~\ref{sec:Generality}, we study in this section anomalies and discrete torsion in the worldsheet theory of Type~II string theory.
Although these issues have already been studied by Delgado, Eberhardt, and Toma\v{s}evi\'{c}~\cite{Delgado:2026qvy}, we revisit them here for two reasons.
First, our discussion of the anomalies in the worldsheet theory is slightly different from that in Ref.~\cite{Delgado:2026qvy}.
The authors of Ref.~\cite{Delgado:2026qvy} directly compute the relevant subgroup of the appropriate bordism group for a general target space by means of the Smith isomorphism.
In contrast, we derive the anomaly cancellation condition by employing the model of the Anderson dual of the bordism theory constructed by Yamashita and Yonekura~\cite{Yamashita:2021cao}.
As reviewed in Sec.~\ref{subsec:anomaly_cancellation}, in this explicit model the anomaly of the worldsheet theory is described as the pullback along the map classifying the tangent bundle of the target space.
Thus, it suffices in practice to compute the bordism group of the classifying space $B\O(10)$.
In this sense, our derivation is new and more general, although the resulting anomaly cancellation condition is of course the same as that obtained in Ref.~\cite{Delgado:2026qvy}.
Second, some anomalies appearing in the worldsheet theory of Type~II string theory also arise in that of Type~I string theory.
Hence, it is useful to examine them here, even for readers interested only in Type~I string theory.

\subsection{Anomalies in Type~II string theory}
\label{subsec:Type_II_anomaly}
In order to discuss anomalies in the worldsheet theory of Type~II string theory, it is necessary to determine the relevant tangential structure and symmetry of the theory.
The worldsheet theory is an $\cN=(1,1)$ nonlinear sigma model with a ten-dimensional target space $X$.
The Majorana fermions in the worldsheet theory are spacetime vectors.
More precisely, they take values in the pullback of the tangent bundle of the target space $X$ along the sigma-model field.
To construct Type~II string theory, separate spin structures are assigned to the left- and right-moving fermions, and all contributions from these assignments are summed over~\cite{Seiberg:1986by,Alvarez-Gaume:1986ghj}.
This prescription is called the chiral GSO projection.\footnote{The term chiral GSO projection is used to distinguish it from the diagonal GSO projection by which Type~0 string theory is formulated.}
In other words, the worldsheet is endowed with a $\Spin\times\bZ_{2}$ structure.
Since the worldsheet of Type~II string theory is oriented, its tangent bundle is an $\SO(n)$-bundle, where $n=2$ for the worldsheet.
In order to assign separate spin structures to the left- and right-moving fermions, the structure group $\SO(n)$ is lifted to a central extension of $\SO(n)$ by the left- and right-moving fermion number symmetries $\bZ_{2}^{L}\times\bZ_{2}^{R}$.
Such a central extension is characterized by an extension class in the group cohomology,
\begin{align}
    H^{2}(B\SO(n);\bZ_{2}^{L}\times\bZ_{2}^{R})
    \cong H^{2}(B\SO(n);\bZ_{2}^{L})\oplus H^{2}(B\SO(n);\bZ_{2}^{R})
    \cong \bZ_{2}\oplus\bZ_{2},
\end{align}
which is generated by the second Stiefel--Whitney class $w_{2}$ in each of the two summands.
Since both $\bZ_{2}^{L}$ and $\bZ_{2}^{R}$ must extend $\SO(n)$ nontrivially, the extension class must be given by $(w_{2},w_{2})\in H^{2}(B\SO(n);\bZ_{2}^{L}\times\bZ_{2}^{R})$.
The group $\Spin(n)\times\bZ_{2}$ fits into the short exact sequence,
\begin{align}\label{eq:extension_II}
    0
    \to \bZ_{2}^{L}\times\bZ_{2}^{R}
    \to \Spin(n)\times\bZ_{2}
    \to \SO(n)
    \to 0,
\end{align}
whose extension class is $(w_{2},w_{2})$.
Intuitively, the $\bZ_{2}$ factor of $\Spin(n)\times\bZ_{2}$ corresponds to the chiral symmetry acting on either the left- or right-moving fermions.

Given these data, the anomaly in the worldsheet theory of Type~II string theory is classified by the group $(I_{\bZ}\Omega^{\Spin\times\bZ_{2}})^{4}(X)$.
Let $f:X\to B\O(10)$ denote the classifying map of the tangent bundle of the target space.
The field content of the worldsheet theory specifies an element $[(h_{0},\omega_{0})]\in(I_{\bZ}\Omega^{\Spin\times\bZ_{2}})^{4}(B\O(10))$.
The anomaly of the worldsheet theory itself is then given by $f^{\ast}[(h_{0},\omega_{0})]\in(I_{\bZ}\Omega^{\Spin\times\bZ_{2}})^{4}(X)$.
In what follows, we study the invertible field theory $[(h_{0},\omega_{0})]$ in detail in order to determine the anomaly cancellation condition for the worldsheet theory.

We begin with the contribution from perturbative anomalies that can be captured by the anomaly polynomial.
There are two types of possible perturbative anomalies in the worldsheet theory.
One is the pure gravitational anomaly of the worldsheet, while the other is the perturbative anomaly of the $\O(10)$ symmetry under which the Majorana fermions belong to the vector representation.
In the present case, the perturbative anomalies are cancelled, since the contributions of the left-moving fermions to the anomaly polynomial have the opposite sign to those of the right-moving fermions.
In other words, both the local Lorentz symmetry of the worldsheet and the $\O(10)$ symmetry are vector-like.
Therefore, we can set $\omega_{0}$ to be zero.

Once the absence of the perturbative anomalies is established, the invertible field theory that captures the global anomalies is regarded as a bordism invariant of $\Omega^{\Spin\times\bZ_{2}}_{3}(B\O(10))$.
Thus, the invertible field theory is characterized by the values of the partition function evaluated on the generators of $\Omega^{\Spin\times\bZ_{2}}_{3}(B\O(10))$, which is a finite group as we will see below.
Before proceeding, we collect several standard facts about bordism groups.
For a connected topological space $Y$, the spin bordism group of $Y$ contains that of a point as a direct summand.
The remaining summand is referred to as the reduced spin bordism group and is denoted by $\tilde{\Omega}^{\Spin}_{n}(Y)$, so that $\Omega^{\Spin}_{n}(Y)\cong\Omega^{\Spin}_{n}(\pt)\oplus\tilde{\Omega}^{\Spin}_{n}(Y)$.
Another fact used repeatedly in this paper is that the bordism group $\Omega^{\Spin\times\bZ_{2}}_{n}(Y)$ is isomorphic to $\Omega^{\Spin}_{n}(B\bZ_{2}\times Y)$.
It is also known that the reduced spin bordism group of a product space $Y_{1}\times Y_{2}$ can be decomposed as
\begin{align}
    \tilde{\Omega}^{\Spin}_{n}(Y_{1}\times Y_{2})
    \cong \tilde{\Omega}^{\Spin}_{n}(Y_{1})
    \oplus \tilde{\Omega}^{\Spin}_{n}(Y_{2})
    \oplus \tilde{\Omega}^{\Spin}_{n}(Y_{1}\wedge Y_{2}).
\end{align}
This decomposition follows from the weak homotopy equivalence,
\begin{align}
    (Y_{1}\times Y_{2})_{+}
    \simeq Y_{1+}\wedge Y_{2+}
    \simeq (S^{0}\vee Y_{1})\wedge(S^{0}\vee Y_{2})
    \simeq S^{0}\vee Y_{1}\vee Y_{2}\vee(Y_{1}\wedge Y_{2}).
\end{align}
Combining these facts, we obtain the following expression,
\begin{align}\label{eq:Type_II_bordism}
    \Omega^{\Spin\times\bZ_2}_3(B\O(10))
    \cong \Omega^{\Spin\times\bZ_{2}}_{3}(\pt)
    \oplus \tilde{\Omega}^{\Spin}_{3}(B\O(10))
    \oplus \tilde{\Omega}^{\Spin}_{3}(B\bZ_{2}\wedge B\O(10)).
\end{align}
As shown in App.~\ref{subsec:Adams_Type_II}, the direct summands appearing in this expression are determined as follows:
\begin{align}\label{eq:Type_II_summand}
    \Omega^{\Spin\times\bZ_{2}}_{3}(\pt)&\cong \bZ_{8},&
    \tilde{\Omega}^{\Spin}_{3}(B\O(10))&\cong \bZ_{8}\oplus\bZ_{2},&
    \tilde{\Omega}^{\Spin}_{3}(B\bZ_{2}\wedge B\O(10))&\cong \bZ_{4}\oplus\bZ_{2}.
\end{align}
In the following, we study the anomalies encoded in the values of the partition function of the invertible field theory on the generators of each direct summand in \eqref{eq:Type_II_bordism}.

Let us discuss the anomaly encoded in the first summand $\Omega^{\Spin\times\bZ_{2}}_{3}(\pt)\cong\bZ_{8}$ in \eqref{eq:Type_II_bordism}.
This anomaly is the obstruction to consistently implementing the chiral GSO projection for Type~II string theory on a flat target space, since $\Omega^{\Spin\times\bZ_{2}}_{3}(\bR^{10})\cong\Omega^{\Spin\times\bZ_{2}}_{3}(\pt)$.
It is known that a generator of $\Omega^{\Spin\times\bZ_{2}}_{3}(\pt)$ is given by the real projective space $\bR\bP^{3}$, whose $\bZ_{2}$-bundle corresponds to the tautological line bundle.
Since a single Majorana--Weyl fermion charged under the chiral $\bZ_{2}$ symmetry contributes $1/8$ to $h_{0}$ evaluated on $\bR\bP^{3}$, the number of physical Majorana fermions in the worldsheet theory must be a multiple of eight.
This condition is indeed satisfied for a ten-dimensional target space~\cite{Kaidi:2019tyf}, either by choosing light-cone gauge or by taking the superconformal ghosts into account.\footnote{
Precisely speaking, there is a subtlety here.
The GSO projection is the procedure of summing over all separate spin structures for the left- and right-moving sectors, whereas what we implement is the gauging of the chiral $\bZ_{2}$ symmetry.
The anomaly relevant to the former takes values in $\bZ_{16}$, whose reduction modulo $8$ is the obstruction to the latter.
Accordingly, when the anomaly vanishes modulo $16$, both operations can be consistently performed, whereas if it equals $8$ modulo $16$, the chiral $\bZ_{2}$ symmetry can still be gauged although the sum over the spin structures no longer yields a bosonic theory.
In this paper, we regard the GSO projection as the gauging of the chiral $\bZ_{2}$ symmetry.
For the worldsheet theory of Type~II string theory, the anomaly is in fact trivial modulo $16$, since the ten worldsheet Majorana fermions contribute $10$ and the superconformal ghosts contribute $22$, which is twice their central charge.
Note that these contributions reduce, modulo $8$, to the counting used in the main text.
See Refs.~\cite{BoyleSmith:2024qgx,Heckman:2025wqd} for a detailed discussion of these subtleties.
}

Next, we focus on the second summand $\tilde{\Omega}^{\Spin}_{3}(B\O(10))\cong\bZ_{8}\oplus\bZ_{2}$ in \eqref{eq:Type_II_bordism}.
In fact, the anomalies encoded in this summand do not arise in the worldsheet theory.
Note that these anomalies must be captured by a worldsheet equipped with the same spin structure for the left- and right-moving fermions, because the bordism group $\tilde{\Omega}^{\Spin}_{3}(B\O(10))$ does not involve the chiral $\bZ_{2}$ symmetry.
On such a worldsheet, there is no phase ambiguity in the fermion path integral, since the fermions can be gapped out by a mass term that breaks the chiral $\bZ_{2}$ symmetry.
As a result, the worldsheet theory does not suffer from the anomalies corresponding to the $\tilde{\Omega}^{\Spin}_{3}(B\O(10))$ summand.

The remaining task is to study the anomaly captured by the last summand $\tilde{\Omega}^{\Spin}_{3}(B\bZ_{2}\wedge B\O(10))\cong\bZ_{4}\oplus\bZ_{2}$ in \eqref{eq:Type_II_bordism}.
These anomalies are encoded in the partition function of the invertible field theory evaluated on a three-dimensional spin manifold $W$ equipped with a real line bundle and a rank-10 real vector bundle.
Note that a cohomology class $\tilde{a}\in H^{1}(W;\bZ_{2})$ determines a real line bundle $L_{\tilde{a}}$ on $W$, and a rank-10 real vector bundle $E$ on $W$ is given by the pullback of the universal bundle along a classifying map $g:W\to B\O(10)$.
The Stiefel--Whitney classes of $E$ are given by $w_{i}(E)=g^{\ast}w_{i}$ for $i=1,2,3$, where $H^{\ast}(B\O(10);\bZ_{2})\cong\bZ_{2}[w_{1},w_{2},\dots,w_{10}]$.
The computation using the Adams spectral sequence in App.~\ref{subsec:Adams_Type_II} suggests that the $\bZ_{4}$ summand is related to $\tilde{a}^{2}w_{1}(E)$ and the $\bZ_{2}$ summand is detected by $\tilde{a}w_{2}(E)$.
The explicit form of the partition function of the invertible field theory is essentially derived in Ref.~\cite{Delgado:2026qvy}.
In order to write down the partition function, let us apply the Witten--Yonekura construction~\cite{Witten:2019bou} to the fermions in the worldsheet theory.
For each of the left- and right-moving Majorana--Weyl fermions, we prepare massive Majorana fermions in three dimensions that yield them as boundary modes.
The Majorana fermions on $W$ for the left-moving fermions are coupled to both the real line bundle $L_{\tilde{a}}$ and the real vector bundle $E$, whereas those for the right-moving fermions are coupled only to the real vector bundle $E$.
Then, the partition function of the invertible field theory that captures the anomalies in the worldsheet theory is expressed in terms of the reduced $\eta$-invariant defined as
\begin{align}\label{eq:reduced_eta}
    \tilde{\eta}(W,L_{\tilde{a}},E)
    :=
    \eta(\cD_{W}^{E\otimes L_{\tilde{a}}})-\eta(\cD_{W}^{E})
    -\left[\eta(\cD_{W}^{L_{\tilde{a}}^{\oplus 10}})-\eta(\cD_{W}^{\cO^{\oplus 10}})\right].
\end{align}
Here, the superscript of each Dirac operator indicates the bundle to which the corresponding massive Majorana fermions couple, and $\cO$ denotes the trivial line bundle.
In the above definition of the reduced $\eta$-invariant, the terms in the square brackets are subtracted in order to remove the part of the anomaly corresponding to $\Omega^{\Spin\times\bZ_{2}}_{3}(\pt)$, which is not relevant for the present purpose.
Recall that the anomalies associated with the first and the second summands in \eqref{eq:Type_II_bordism} were shown to be absent above, and that $\tilde{\eta}$ vanishes identically on these two summands by construction.
Thus, the invertible field theory is completely determined by $\tilde{\eta}$.
The partition function is then determined in Ref.~\cite{Delgado:2026qvy}\footnote{In Ref.~\cite{Delgado:2026qvy}, the partition function is determined as an element of $\Hom(\Omega^{\Spin}_{3}(B\bZ_{2}\wedge Y),\bR/\bZ)$ for a general target space $Y$. For our purposes, it suffices to use the special case of this formula for $Y=B\O(10)$.} as
\begin{align}\label{eq:Type_II_IFT}
    e^{2\pi\i h_{0}(W,\tilde{a},g)}=e^{-\pi\i\tilde{\eta}(W,L_{\tilde{a}},E)}=\exp\left(\pi\i\int_{W}\tilde{a}w_{2}(E)\right)\exp\left(\pm\frac{\pi\i}{2}q_{S}(w_{1}(E|_{S}))\right).
\end{align}
The $\bZ_{4}$-valued quantity $q_{S}(w_{1}(E|_{S}))$ is defined as follows.
Let $S\subset W$ be a surface Poincar\'e dual to the class $\tilde{a}\in H^{1}(W;\bZ_{2})$.\footnote{The surface $S$ is equipped with the $\Pin^{-}$ structure determined by the $\Spin\times\bZ_{2}$ structure on $W$.}
On the surface $S$, there is a quadratic enhancement $q_{S}:H^{1}(S;\bZ_{2})\to\bZ_{4}$ such that
\begin{align}
    q_{S}(s_{1}+s_{2})-q_{S}(s_{1})-q_{S}(s_{2})=2\int_{S}\,s_{1}\smile s_{2},
\end{align}
for $s_{1}, s_{2}\in H^{1}(S;\bZ_{2})$.
The quantity $q_{S}(w_{1}(E|_{S}))$ is the value of this quadratic enhancement on the first Stiefel--Whitney class of the restriction $E|_{S}$.

Anomaly cancellation requires that $f^{\ast}h_{0}=0$ on the torsion subgroup of $\Omega^{\Spin\times\bZ_{2}}_{3}(X)$.
To examine this condition, let us take a bordism class of $\Omega^{\Spin\times\bZ_{2}}_{3}(X)$ represented by $(W,\tilde{a},\phi)$, where $(W,\tilde{a})$ is a three-manifold equipped with a $\Spin\times\bZ_{2}$ structure and $\phi$ is a map from $W$ to $X$.
The real vector bundle on $W$ is obtained as the pullback $\phi^{\ast}TX$ of the tangent bundle of $X$.
We then have
\begin{align}\begin{aligned}\label{eq:Type_II_anomaly}
    e^{2\pi\i f^{\ast}h_{0}(W,\tilde{a},\phi)}
    &=\exp\left(-\pi\i \tilde{\eta}(W,L_{\tilde{a}},\phi^{\ast}TX)\right)\\
    &=\exp\left(\pi\i\int_{W}\tilde{a}\smile\phi^{\ast}w_{2}(TX)\right)\exp\left(\pm\frac{\pi\i}{2}q_{S}(w_{1}(\phi^{\ast}TX|_{S}))\right).
\end{aligned}\end{align}
This phase is trivial if both $w_{1}(TX)=f^{\ast}w_{1}$ and $w_{2}(TX)=f^{\ast}w_{2}$ vanish.
In other words, the anomalies in the worldsheet theory are cancelled if the target space is orientable and admits a spin structure.

Conversely, for the anomalies to be cancelled, the target space must be orientable and admit a spin structure.
To see this, it is helpful to understand the physical origin of the anomalies described by the invertible phase~\eqref{eq:Type_II_anomaly}.
In order to discuss the anomaly involving the first Stiefel--Whitney class, let us consider a configuration of the sigma-model field in which the worldsheet is mapped to a point $x\in X$.
We also assume that the worldsheet is equipped with different spin structures for the left- and right-moving fermions, so that the number of zero modes is odd for each left-moving fermion and even for each right-moving fermion.\footnote{The argument presented here can be found, e.g., in Ref.~\cite{Witten:2023snr}.}
For simplicity, we assume that each Majorana fermion produces a single zero mode denoted by $\psi^{i}_{0}$.
In this situation, the measure of the fermion path integral includes $\d\psi^{i}_{0}$ for $i=0,1,\dots,9$.
Since there is no canonical choice of the ordering of these factors, we need to choose one by hand.
However, if the point $x\in X$ is transported along an orientation-reversing loop in $X$, the chosen ordering is reversed.
The fermion path integral is then ill-defined for this configuration.
In order to avoid this inconsistency, the target space $X$ must be orientable.

Similarly, the physical origin of the other anomaly involving the second Stiefel--Whitney class in the invertible phase~\eqref{eq:Type_II_anomaly} can be understood by considering the following situation.
We assume that the worldsheet is a torus and that the target space is oriented.\footnote{As we will mention in Sec.~\ref{subsec:Type_II_discrete_torsion}, the target space is not only orientable but also oriented.}
Along a spatial cycle, the periodic boundary condition is imposed on the left-moving fermions, whereas the antiperiodic boundary condition is imposed on the right-moving fermions; that is, we consider the RNS sector.
Suppose that the image of the torus under the sigma-model map is topologically a circle in the target space, and that the temporal cycle of the torus wraps a nontrivial cycle of $X$.
In the RNS sector, the spatial zero modes arise only from the left-moving fermions, and they can be regarded as one-dimensional Majorana fermions.
Since the target space is now oriented, these Majorana fermions are coupled to an $\SO(10)$-bundle on the temporal circle, which is the pullback of the tangent bundle of $X$.
This system of one-dimensional Majorana fermions suffers from a global anomaly~\cite{Yonekura:2022reu} by an argument similar to that for the Witten anomaly~\cite{Witten:1982fp}.
In this case, the anomaly is captured by $\pi_{1}(\SO(10))\cong\bZ_{2}$, and is nothing but the anomaly associated with the second Stiefel--Whitney class in \eqref{eq:Type_II_anomaly}.
To cancel this anomaly, the target space must admit a spin structure.\footnote{By construction, these anomalies are detected by mapping tori $W$ whose fundamental classes have trivial images in $H_{3}(X;\bZ)$. Since $\Omega^{\Spin\times\bZ_{2}}_{3}(X)\otimes\bQ\cong H_{3}(X;\bQ)$, the corresponding bordism classes are torsion elements of $\Omega^{\Spin\times\bZ_{2}}_{3}(X)$, and hence neither of the anomalies discussed in this paragraph and in the previous one can be absorbed by a local counterterm.}
It is worth emphasizing that this requirement, which may look natural from the spacetime point of view, follows purely from the consistency of the worldsheet theory.

\subsection{Discrete torsion of Type~II string theory}
\label{subsec:Type_II_discrete_torsion}

We next discuss the discrete torsion in the worldsheet theory of Type~II string theory.
Throughout this subsection, we assume that the target space $X$ is orientable and admits a spin structure, so that the anomalies discussed in Sec.~\ref{subsec:Type_II_anomaly} are cancelled.
As mentioned in Sec.~\ref{subsec:Yamashita-Yonekura}, the discrete torsion is part of the data defining the worldsheet theory, and any two choices of it differ by an element of $\Hom(\Omega^{\Spin\times\bZ_{2}}_{2}(X),\bR/\bZ)$.
The complete analysis of the discrete torsion for Type~II string theory has been carried out in Ref.~\cite{Delgado:2026qvy}.
Here, we summarize the results, since some of them also appear in Type~I string theory.

To this end, it is necessary to determine the bordism group $\Omega^{\Spin\times\bZ_{2}}_{2}(X)$.
As in Sec.~\ref{subsec:Type_II_anomaly}, this bordism group can be decomposed as
\begin{align}
\label{eq:Type_II_discrete_torsion}
    \Omega^{\Spin\times\bZ_{2}}_{2}(X)
    \cong \Omega^{\Spin\times\bZ_{2}}_{2}(\pt)
    \oplus \tilde{\Omega}^{\Spin}_{2}(X)
    \oplus \tilde{\Omega}^{\Spin}_{2}(B\bZ_{2}\wedge X).
\end{align}
Each summand in this decomposition can be computed by means of the Atiyah--Hirzebruch spectral sequence (AHSS).\footnote{See, e.g., Ref.~\cite{Garcia-Etxebarria:2018ajm} for a review of the AHSS aimed at physicists.}
It follows from the AHSS, together with $H_{1}(B\bZ_{2};\bZ_{2})\cong\bZ_{2}$ and $H_{2}(B\bZ_{2};\bZ)=0$, that the first summand is given by
\begin{align}
    \Omega^{\Spin\times\bZ_{2}}_{2}(\pt)
    \cong \Omega^{\Spin}_{2}(\pt) \oplus H_{1}(B\bZ_{2};\bZ_{2})
    \cong \bZ_{2}\oplus\bZ_{2}.
\end{align}
In the same way, the second summand fits into the following short exact sequence:
\begin{align}\label{eq:Type_II_second_torsion}
    0
    \to H_{1}(X;\bZ_{2})
    \to \tilde{\Omega}^{\Spin}_{2}(X)
    \to H_{2}(X;\bZ)
    \to 0.
\end{align}
By means of the Smith isomorphism,\footnote{See, e.g., Refs.~\cite{Tachikawa:2018njr,Hason:2020yqf} for an introduction to the Smith isomorphism in the context of high-energy physics.} the last summand is computed as
\begin{align}
    \tilde{\Omega}^{\Spin}_{2}(B\bZ_{2}\wedge X)
    \cong \tilde{\Omega}^{\Pin^{-}}_{1}(X)
    \cong H_{1}(X;\bZ_{2}).
\end{align}
Here, we have again applied the AHSS, now converging to the $\Pin^{-}$ bordism group, together with $\Omega^{\Pin^{-}}_{0}(\pt)\cong\bZ_{2}$.

The discrete torsion captured by $\Hom(\Omega^{\Spin\times\bZ_{2}}_{2}(\pt),\bR/\bZ)\cong\bZ_{2}\oplus\bZ_{2}$ is generated by the Arf invariants of the spin structures for the left- and right-moving sectors.
As explained in Ref.~\cite{Kaidi:2019tyf}, the summand $\Hom(\Omega^{\Spin}_{2}(\pt),\bR/\bZ)$ of this group corresponds to the distinction between Type~IIA and Type~IIB string theories, whereas the other summand represents the possible choices of the orientation of the target space.

For an intuitive understanding of the remaining summands, let us first assume that the sequence~\eqref{eq:Type_II_second_torsion} splits.
Then, there are two $H_{1}(X;\bZ_{2})$ summands in \eqref{eq:Type_II_discrete_torsion}, and the corresponding discrete torsions lie in $\Hom(H_{1}(X;\bZ_{2}),\bR/\bZ)\cong H^{1}(X;\bZ_{2})$.
They are generated by the quadratic refinements associated with the spin structures for the left- and right-moving fermions.
Recall that there are two $\bZ_{2}$ symmetries~\cite{Tachikawa:2018njr,Heckman:2025wqd} arising as the quantum symmetries~\cite{Vafa:1989ih} associated with the gauging of the left- and right-moving fermion number symmetries.
Their diagonal combination is nothing but the spacetime fermion number symmetry.
Accordingly, the summand $\tilde{\Omega}^{\Spin}_{2}(B\bZ_{2}\wedge X)$ is the one detected by this diagonal symmetry, and it parametrizes the possible shifts of the spin structure of the target space from a fixed reference one.
On the other hand, the remaining $H^{1}(X;\bZ_{2})$ summand corresponds to a background gauge field on $X$ for either of the two quantum symmetries~\cite{Delgado:2026qvy}.
As we will see, these types of discrete torsion, characterized by $H^{1}(X;\bZ_{2})$, also appear in the worldsheet theory of Type~I string theory.
The discrete torsion associated with $\Hom(H_{2}(X;\bZ),\bR/\bZ)\cong H^{2}(X;\bR/\bZ)$ is the flat part of the $B$-field on the target space~\cite{Witten:1999eg}.
When the sequence~\eqref{eq:Type_II_second_torsion} does not split, we have the exact sequence of the Pontryagin dual
\begin{align}
    0\to H^{2}(X;\bR/\bZ)\to\Hom(\tilde{\Omega}^{\Spin}_{2}(X),\bR/\bZ)\to H^{1}(X;\bZ_{2})\to0.
\end{align}
Two of these discrete torsions reproduce a flat $B$-field, which is given by their cup product.
In the context of condensed matter physics, this is known as the Gu--Wen extension~\cite{Gu:2012ib,Bhardwaj:2016clt}.

\section{Type I string theory}
\label{sec:Type_I}

In this section, we study anomalies and discrete torsion in the worldsheet theory of Type~I string theory.
The worldsheet of Type~I string theory is equipped with a $\DPin$ structure.
The group $\DPin(n)$, which was introduced in Ref.~\cite{Kaidi:2019tyf}, is reviewed in Sec.~\ref{subsec:DPin}.
The anomaly cancellation condition and the discrete torsion are discussed in Sec.~\ref{subsec:Type_I_anomaly} and Sec.~\ref{subsec:Type_I_discrete_torsion}, respectively.

\subsection{\texorpdfstring{$\mathrm{DPin}$}{DPin} structure}
\label{subsec:DPin}
In order to obtain Type~I string theory, the chiral GSO projection is combined with the gauging of the worldsheet orientation reversal.
This operation is referred to as the orientifold projection.
As a result, the worldsheet theory is defined not only on orientable worldsheets but also on unorientable ones.
The structure group $\O(n)$ of the tangent bundle of the worldsheet is lifted to $\DPin(n)$~\cite{Kaidi:2019tyf}, which fits into the short exact sequence:
\begin{equation}\label{eq:extension_I-1}
    0
    \to \bZ_{2}^{L}\times\bZ_{2}^{R}
    \to \DPin(n)
    \to \O(n)
    \to 0.
\end{equation}
Here, $\bZ_{2}^{L}$ and $\bZ_{2}^{R}$ correspond to the fermion number symmetries associated with the left- and right-moving sectors, respectively.
The group $\O(n)$ acts on $\bZ_{2}^{L}\times\bZ_{2}^{R}$ through a homomorphism $\rho:\O(n)\to\mathrm{Aut}(\bZ_{2}^{L}\times\bZ_{2}^{R})$, under which an element of determinant $-1$ exchanges the two factors.
The origin of this action $\rho$ is the worldsheet orientation reversal, which exchanges the left- and right-moving sectors.
Accordingly, when the worldsheet fermions are transported along an orientation-reversing loop, the left- and right-moving fermion number symmetries are interchanged, and neither of them is separately well-defined on an unoriented worldsheet.
The group $\DPin(n)$ is defined by specifying the extension class in $H^2(B\O(n);\bZ_{2}^{L}\times\bZ_{2}^{R}|_{\rho})$.
Since the sequence~\eqref{eq:extension_I-1} is not a central extension, the coefficient system appearing in this classification is twisted by $\rho$, which makes it inconvenient to express the extension class in terms of characteristic classes.
We therefore introduce the homomorphism $s:\bZ_{2}^{L}\times\bZ_{2}^{R}\to\bZ_{2}^{\mathrm{(rel)}}$ defined by $s(\alpha_{1},\alpha_{2})=\alpha_{1}\alpha_{2}$, where $\bZ_{2}^{L}\times\bZ_{2}^{R}$ is written multiplicatively, and denote its kernel by $\bZ_{2}^{\mathrm{(diag)}}=\Ker s$.
Since the diagonal subgroup $\bZ_{2}^{\mathrm{(diag)}}\subset\bZ_{2}^{L}\times\bZ_{2}^{R}$ is invariant under the action $\rho$, it lies in the center of $\DPin(n)$.
This subgroup is nothing but the total fermion number symmetry.
Consequently, the extensions considered below are classified by cohomology groups with untwisted coefficients.
We then obtain the short exact sequence:
\begin{equation}\label{eq:extension_I-2}
    0
    \to \bZ_{2}^{\mathrm{(rel)}}
    \to \frac{\DPin(n)}{\bZ_{2}^{\mathrm{(diag)}}}
    \to \O(n)
    \to 0.
\end{equation}
Note that the group $\O(n)$ acts trivially on $\bZ_{2}^{\mathrm{(rel)}}$.
As shown in Ref.~\cite{Kaidi:2019tyf}, the group extension~\eqref{eq:extension_I-2} is trivial, and the possible identifications $\DPin(n)/\bZ_{2}^{\mathrm{(diag)}}\cong\O(n)\times\bZ_{2}^{\mathrm{(rel)}}$ are parametrized by $H^{1}(B\O(n);\bZ_{2}^{\mathrm{(rel)}})\cong\bZ_{2}$.
This is consistent with the fact that, upon restriction to $\SO(n)\subset\O(n)$, the sequence~\eqref{eq:extension_I-1} reduces to the sequence~\eqref{eq:extension_II}, so that the extension class of the sequence~\eqref{eq:extension_I-2} becomes $w_{2}+w_{2}=0$.
Given an identification, we obtain the short exact sequence:
\begin{align}
    0
    \to \bZ_{2}^{\mathrm{(diag)}}
    \to \DPin(n)
    \to \O(n)\times\bZ_{2}^{\mathrm{(rel)}}
    \to 0.
\end{align}
This sequence is a central extension, and is therefore classified by an element of $H^{2}(B\O(n)\times B\bZ_{2}^{\mathrm{(rel)}};\bZ_{2}^{\mathrm{(diag)}})$.
In the present case, the extension class is given by $w_{2}+w_{1}^{2}+w_{1}a$, where $w_{i}$ are the Stiefel--Whitney classes of the universal $\O(n)$-bundle and $a$ is the nontrivial element of $H^{1}(B\bZ_{2}^{\mathrm{(rel)}};\bZ_{2}^{\mathrm{(diag)}})$.
Equivalently, a $\DPin$ structure on a manifold $M$ consists of a class $a\in H^{1}(M;\bZ_{2})$, which specifies the $\bZ_{2}^{\mathrm{(rel)}}$-bundle, together with a trivialization of $w_{2}(TM)+w_{1}(TM)^{2}+w_{1}(TM)a$.
The two identifications labeled by $H^{1}(B\O(n);\bZ_{2}^{\mathrm{(rel)}})$ are related by the replacement $a\to w:=a+w_{1}$.
In particular, setting $a=0$ gives the condition $w_{2}+w_{1}^{2}=0$ for a $\Pin^{-}$ structure and setting $w=0$ gives $w_{2}=0$ for a $\Pin^{+}$ structure.
Restricting to oriented manifolds imposes $w_1=0$ and leaves the condition $w_{2}=0$ with $a$ unconstrained, which is nothing but a $\Spin\times\bZ_{2}$ structure.
Accordingly, the group $\DPin(n)$ contains $\Spin(n)\times\bZ_{2}$, $\Pin^{+}(n)$, and $\Pin^{-}(n)$ as subgroups.

\subsection{Anomalies in Type~I string theory}
\label{subsec:Type_I_anomaly}

We now consider the worldsheet theory of Type~I string theory, whose target space is a ten-dimensional manifold $X$.
Let $f:X\to B\O(10)$ denote the classifying map of the tangent bundle of $X$.
Based on the general argument reviewed in Sec.~\ref{sec:Generality}, the anomaly in the worldsheet theory is classified by the group $(I_{\bZ}\Omega^{\DPin})^{4}(X)$.\footnote{In this paper, we study only the anomalies in the closed string sector. The analysis of the open string sector is left for future work.}
The field content of the worldsheet theory specifies an element $[(h_{0},\omega_{0})]\in(I_{\bZ}\Omega^{\DPin})^{4}(B\O(10))$.
The anomaly of the worldsheet theory itself is then given by $f^{\ast}[(h_{0},\omega_{0})]\in(I_{\bZ}\Omega^{\DPin})^{4}(X)$.
In what follows, we study the invertible field theory $[(h_{0},\omega_{0})]$ in order to determine the anomaly cancellation condition.

The perturbative anomalies are absent for the same reason as in Type~II string theory.
The possible perturbative anomalies are the pure gravitational anomaly of the worldsheet and the anomaly of the $\O(10)$ symmetry.
Since both the local Lorentz symmetry of the worldsheet and the $\O(10)$ symmetry are vector-like, the contributions of the two chiralities to the anomaly polynomial cancel each other.
Here, the two chiralities are distinguished only locally, as discussed in Sec.~\ref{subsec:DPin}; this is nevertheless sufficient, because the anomaly polynomial is a local quantity.
Therefore, we can set $\omega_{0}$ to be zero.

The invertible field theory corresponding to $[(h_{0},0)]$ is characterized by the values of the partition function evaluated on the generators of the torsion subgroup of $\Omega^{\DPin}_{3}(B\O(10))$.
Since $\Omega^{\DPin}_{n}(\pt)$ is a torsion group for $n=0,1,2,3$, it follows from the Atiyah--Hirzebruch spectral sequence that $\Omega^{\DPin}_{n}(X)$ is a torsion group for any target space $X$ and any $n\leq3$.
The anomaly cancellation condition is therefore simply the vanishing of $f^{\ast}h_{0}$ on all of $\Omega^{\DPin}_{3}(X)$.
As shown in App.~\ref{subsec:Adams_Type_I}, the bordism group $\Omega^{\DPin}_{3}(B\O(10))$ is determined as
\begin{align}\label{eq:Type_I_bordism}
    \Omega^{\DPin}_{3}(B\O(10))
    \cong \Omega^{\DPin}_{3}(\pt)
    \oplus \tilde{\Omega}^{\Spin}_{3}(B\bZ_{2}\wedge B\O(10))
    \oplus \bZ_{2}^{\oplus 4},
\end{align}
where
\begin{align}
    \Omega^{\DPin}_{3}(\pt)\cong\bZ_{8},\quad
    \tilde{\Omega}^{\Spin}_{3}(B\bZ_{2}\wedge B\O(10))\cong\bZ_{4}\oplus\bZ_{2}.
\end{align}
In order to write the second summand in this form, we have used the fact that the bordism group $\Omega^{\DPin}_{3}(B\O(10))$ contains $\tilde{\Omega}^{\Spin}_{3}(B\bZ_{2}\wedge B\O(10))$ as a direct summand, as shown in Proposition~5.107 of Ref.~\cite{Debray:2026ivi}.
In the following, we discuss the anomalies encoded in each summand in turn.

Let us first discuss the anomaly encoded in $\Omega^{\DPin}_{3}(\pt)\cong\bZ_{8}$.
As discussed in Ref.~\cite{Kaidi:2019tyf}, this anomaly is cancelled for a ten-dimensional target space, just as in Type~II string theory.

Next, we move on to the anomalies encoded in $\tilde{\Omega}^{\Spin}_{3}(B\bZ_{2}\wedge B\O(10))$.
Since the group $\DPin(n)$ contains $\Spin(n)\times\bZ_{2}$ as a subgroup, a manifold with a $\Spin\times\bZ_{2}$ structure can also be regarded as one with a $\DPin$ structure.
Hence, this direct summand is realized as the image of the map $\Omega^{\Spin\times\bZ_{2}}_{3}(B\O(10))\to\Omega^{\DPin}_{3}(B\O(10))$ induced by this inclusion.
Recall that $\tilde{\Omega}^{\Spin}_{3}(B\bZ_{2}\wedge B\O(10))$ is also a direct summand of $\Omega^{\Spin\times\bZ_{2}}_{3}(B\O(10))$.
Moreover, on an orientable worldsheet, the field content of Type~I string theory is the same as that of Type~II string theory.
Thus, the anomalies encoded in $\tilde{\Omega}^{\Spin}_{3}(B\bZ_{2}\wedge B\O(10))$ are captured on three-manifolds equipped with a $\Spin\times\bZ_{2}$ structure, and they coincide with those analyzed in Sec.~\ref{subsec:Type_II_anomaly}.
These anomalies are cancelled if and only if the target space is orientable and admits a spin structure.

It remains to study the four $\bZ_{2}$ summands in \eqref{eq:Type_I_bordism}.
The anomalies corresponding to these summands are encoded in the partition function of the invertible field theory evaluated on a three-dimensional manifold $W$ equipped with a $\DPin$ structure and a rank-10 real vector bundle.
As reviewed in Sec.~\ref{subsec:DPin}, a cohomology class $\tilde{a}\in H^{1}(W;\bZ_{2})$ is part of the data specifying a $\DPin$ structure.
For later use, we introduce the notation $\tilde{w}:=\tilde{a}+w_{1}(TW)$.
A real vector bundle $E$ on $W$ associated with the $\O(10)$ symmetry is given by the pullback of the universal bundle along a classifying map $g:W\to B\O(10)$.
As shown in App.~\ref{subsec:Adams_Type_I}, the four $\bZ_{2}$ summands are captured by $\tilde{w}^{2}w_{1}(E)$, $w_{1}(E)^{3}$, $w_{1}(E)w_{2}(E)$, and $w_{3}(E)$.

Recall that the analysis of the summand $\tilde{\Omega}^{\Spin}_{3}(B\bZ_{2}\wedge B\O(10))$ already requires $w_{1}(TX)=w_{2}(TX)=0$.
For a configuration of the sigma model, the real vector bundle on $W$ is obtained as the pullback $E=\phi^{\ast}TX$ of the tangent bundle of $X$ along a map $\phi:W\to X$.
We then have $w_{1}(E)=w_{2}(E)=0$, and the Wu formula $\Sq^{1}w_{2}(E)=w_{1}(E)w_{2}(E)+w_{3}(E)$ gives $w_{3}(E)=0$ as well.
Hence, the four characteristic numbers listed above all vanish identically once the target space is orientable and admits a spin structure.
Therefore, these summands impose no further condition on the target space.

In order to understand the physical meaning of the anomaly captured by $\tilde{w}^{2}w_{1}(E)$, we consider the worldsheet theory on the real projective  plane $\bR\bP^{2}$.
Suppose that the sigma-model field maps $\bR\bP^{2}$ to a point $x\in X$.
Note that $\bR\bP^{2}$ generates $\Omega^{\Pin^{-}}_{2}(\pt)\cong\bZ_{8}$, the isomorphism being given by the Arf--Brown--Kervaire invariant~\cite{MR0293642}.
For the $\DPin$ structure on $\bR\bP^{2}$ with $\tilde{a}=0$, which reduces to a $\Pin^{-}$ structure, the Arf--Brown--Kervaire invariant is odd.
Correspondingly, the mod-two index of the Dirac operator is nontrivial.
Hence, each of the ten Majorana fermions has a zero mode in this configuration.
When the point $x\in X$ is transported along an orientation-reversing loop in $X$, the sign of the measure of these zero modes is reversed, by the same argument as in Sec.~\ref{subsec:Type_II_anomaly}.
The fermion path integral is then ill-defined for this configuration.
This inconsistency is avoided once the target space $X$ is orientable, which has already been required above.
Therefore, the present analysis does not impose a new condition.
The purpose of this paragraph is rather to clarify the physical origin of the anomaly captured by $\tilde{w}^{2}w_{1}(E)$, which is specific to Type~I string theory.

In summary, the anomaly cancellation condition for the worldsheet theory of Type~I string theory is that the target space $X$ is orientable and admits a spin structure.

\subsection{Discrete torsion of Type~I string theory}
\label{subsec:Type_I_discrete_torsion}

We finally discuss the discrete torsion in the worldsheet theory of Type~I string theory.
As in Sec.~\ref{subsec:Type_I_anomaly}, we restrict ourselves to the closed string sector.
Throughout this subsection, we assume that the ten-dimensional target space $X$ is orientable and admits a spin structure, so that the anomaly discussed in Sec.~\ref{subsec:Type_I_anomaly} is cancelled.
As explained in Sec.~\ref{subsec:Yamashita-Yonekura}, the possible choices of the discrete torsion form a torsor over $\Hom(\Omega^{\DPin}_{2}(X),\bR/\bZ)$.
For a connected target space $X$, it follows from Proposition~5.107 of Ref.~\cite{Debray:2026ivi} that
\begin{align}\label{eq:discrete_torsion_I}
    \Omega^{\DPin}_{2}(X)
    &\cong \Omega^{\DPin}_{2}(\pt)
    \oplus \tilde{\Omega}^{\Spin}_{2}(B\bZ_{2}\wedge X)
    \oplus H_{2}(X;\bZ_{2}).
\end{align}

It is known that $\Omega^{\DPin}_{2}(\pt)\cong\bZ_{2}\oplus\bZ_{2}$~\cite{Kaidi:2019tyf}.
The topological terms associated with the Pontryagin dual $\Hom(\Omega^{\DPin}_{2}(\pt),\bR/\bZ)\cong\bZ_{2}\oplus\bZ_{2}$ are the Arf invariant of the orientation double cover of the worldsheet $\Sigma$ and the characteristic number $w_{1}(T\Sigma)^{2}$.
When the worldsheet is orientable, the two sheets of its orientation double cover correspond to the left- and right-moving sectors, since the action $\rho$ exchanges $\bZ_{2}^{L}$ and $\bZ_{2}^{R}$ under orientation reversal.
The spin structure induced by the $\DPin$ structure therefore restricts on each sheet to the spin structure assigned to the corresponding sector, and the Arf invariant above reduces to the sum of the Arf invariants of these two spin structures.
As in Type~II string theory, coupling this topological term amounts to flipping the orientation of the target space.

On the other hand, the topological term $w_{1}(T\Sigma)^{2}$ distinguishes ordinary Type~I string theory from Sugimoto string theory~\cite{Sugimoto:1999tx}.
Although we restrict ourselves to the closed string sector, it is worth commenting on the role of this term in the presence of boundaries.
While $w_{1}(T\Sigma)^{2}$ is well-defined on a closed worldsheet, it gives rise to an anomaly once the worldsheet has a boundary.
In order to cancel this anomaly, the Chan--Paton degrees of freedom~\cite{Paton:1969je} on the D$9$-branes must produce a compensating anomaly.
As a result, the gauge group is $\USp(32)$ in the presence of this topological term and $\SO(32)$ in its absence.\footnote{The precise gauge group of ordinary Type~I string theory is given by $\Spin(32)/\bZ_{2}$~\cite{Polchinski:1995df,Witten:1998cd}, while that of Sugimoto string theory is $\USp(32)/\bZ_{2}$~\cite{Larotonda:2024thv}.}
See Refs.~\cite{Kaidi:2019tyf,Witten:2023snr} for more details.

The discrete torsion parametrized by $\Hom(\tilde{\Omega}^{\Spin}_{2}(B\bZ_{2}\wedge X),\bR/\bZ)\cong H^{1}(X;\bZ_{2})$ is one of those already encountered in Type~II string theory.
Recall that Type~II string theory has two discrete torsions of this type, associated with the quantum symmetries arising from the gauging of the left- and right-moving fermion number symmetries.
The orientifold projection exchanges these two symmetries, so that only their diagonal combination survives.
Coupling the corresponding topological term shifts the spin structure of the target space.

The last summand in \eqref{eq:discrete_torsion_I} yields the discrete torsion labeled by $\Hom(H_{2}(X;\bZ_{2}),\bR/\bZ)\cong H^{2}(X;\bZ_{2})$.
This discrete torsion is the flat part of the $B$-field on the target space.
Although the orientifold projection removes the $B$-field from the massless spectrum, a $\bZ_{2}$-valued degree of freedom survives~\cite{Bianchi:1991eu,Sen:1997pm,Witten:1997bs}.
In terms of bordism groups, this is reflected in the appearance of $H_{2}(X;\bZ_{2})$ rather than $H_{2}(X;\bZ)$ in \eqref{eq:discrete_torsion_I}.
The origin of this difference is $\Omega^{\DPin}_{0}(\pt)\cong\bZ_{2}$ instead of $\Omega^{\Spin}_{0}(\pt)\cong\bZ$.
This is the same property that makes $\Omega^{\DPin}_{n}(X)$ a torsion group, a fact we used in Sec.~\ref{subsec:Type_I_anomaly}.

Let us close this subsection by summarizing how the discrete torsion of Type~I string theory differs from that of Type~II string theory.
Among the discrete torsions labeled by the Pontryagin dual $\Hom(\Omega^{\DPin}_{2}(\pt),\bR/\bZ)$, the one that flips the orientation of the target space is still present.
On the other hand, the one distinguishing Type~IIA from Type~IIB string theory is replaced by the topological term $w_{1}(T\Sigma)^{2}$, which instead distinguishes ordinary Type~I string theory from Sugimoto string theory.
The discrete torsions labeled by $H^{1}(X;\bZ_{2})$ are reduced from two to one, since the orientifold projection identifies the quantum symmetries for the left- and right-moving sectors.
Finally, the flat part of the $B$-field is reduced from $H^{2}(X;\bR/\bZ)$ to $H^{2}(X;\bZ_{2})$.
Both of the last two reductions originate from the orientifold projection, which relates the two chiralities on the worldsheet.

\section{Discussions}
\label{sec:Discussion}

Let us first comment on Type~0 string theory, which is defined by the diagonal GSO projection.
In this case, the worldsheet carries a $\Spin$ structure alone, and the anomaly of the worldsheet theory is classified by $(I_{\bZ}\Omega^{\Spin})^{4}(X)$.
As in Sec.~\ref{subsec:anomaly_cancellation}, the worldsheet fermions couple to the target space only through the pullback of its tangent bundle, so that the anomaly again lies in the image of $f^{\ast}:(I_{\bZ}\Omega^{\Spin})^{4}(B\O(10))\to(I_{\bZ}\Omega^{\Spin})^{4}(X)$.
The perturbative anomaly is absent for the same reason as in Type~II and Type~I string theories.
The global anomaly is therefore captured by the values of the partition function of the invertible field theory on the generators of $\Omega^{\Spin}_{3}(B\O(10))$, which has already appeared in \eqref{eq:Type_II_summand}.
Since $\Omega^{\Spin}_{3}(\pt)=0$, this group is given by $\tilde{\Omega}^{\Spin}_{3}(B\O(10))\cong\bZ_{8}\oplus\bZ_{2}$.
This is precisely the summand whose anomaly was shown to be absent in Sec.~\ref{subsec:Type_II_anomaly}, by the argument that the fermions on a worldsheet equipped with a common spin structure can be gapped out by a mass term.
We therefore conclude that the worldsheet theory of Type~0 string theory has no anomaly, and that no additional topological condition is imposed on the target space, in agreement with the statement in Ref.~\cite{Delgado:2026qvy}.

The discrete torsion of Type~0 string theory is classified by $\Hom(\Omega^{\Spin}_{2}(X),\bR/\bZ)$.
It decomposes as $\Omega^{\Spin}_{2}(X)\cong\Omega^{\Spin}_{2}(\pt)\oplus\tilde{\Omega}^{\Spin}_{2}(X)$ with $\Omega^{\Spin}_{2}(\pt)\cong\bZ_{2}$, and the reduced part $\tilde{\Omega}^{\Spin}_{2}(X)$ fits into the short exact sequence~\eqref{eq:Type_II_second_torsion}.
The topological term associated with the Pontryagin dual $\Hom(\Omega^{\Spin}_{2}(\pt),\bR/\bZ)\cong\bZ_{2}$ is the Arf invariant of the spin structure of the worldsheet, and coupling it distinguishes Type~0A from Type~0B string theory~\cite{Kaidi:2019tyf}.
If the sequence~\eqref{eq:Type_II_second_torsion} splits, $H_{2}(X;\bZ)$ contributes $\Hom(H_{2}(X;\bZ),\bR/\bZ)\cong H^{2}(X;\bR/\bZ)$, which is the flat part of the $B$-field.
On the other hand, the subgroup $H_{1}(X;\bZ_{2})$ contributes $H^{1}(X;\bZ_{2})$, which is the background gauge field for the quantum $\bZ_{2}$ symmetry arising from the gauging of the diagonal fermion number symmetry.
It should be emphasized that, in contrast to Type~II string theory, there is no summand corresponding to a shift of the orientation or the spin structure of the target space.
This is consistent with the fact that the target space is not required to be orientable or to admit a spin structure.
As in Type~II string theory, the sequence~\eqref{eq:Type_II_second_torsion} need not split, in which case the discrete torsion forms an extension of $H^{1}(X;\bZ_{2})$ by $H^{2}(X;\bR/\bZ)$.

In this paper, we have focused on the closed string sector of Type~I string theory.
For Type~I string theory to be consistent, however, the open string sector must be introduced to cancel the tadpoles.
Extending our analysis in this direction requires a modern formulation of anomalies for field theories on manifolds with boundary.
An earlier analysis of anomalies in the presence of boundaries was carried out in Refs.~\cite{Witten:1998cd,Freed:1999vc,Kapustin:1999di,Gao:2010ava}.
More recently, Refs.~\cite{Kaidi:2019tyf,Witten:2023snr} developed a treatment based on bordism theory and clarified the relation between the anomaly and the K-theoretic classification of D-branes.
It would be interesting to extend the analysis to the open string sector and to determine the anomaly and the discrete torsion of Type~I string theory on a general target space.

It would also be interesting to generalize the analysis in this paper to other string theories, such as orbifold and orientifold theories.
Orbifolds of Type~II string theory were already discussed in Ref.~\cite{Delgado:2026qvy}.
In general, when a group $\Gamma$ acts nontrivially on the target space $X$, one must consider the bordism groups of the Borel construction, $E\Gamma\times_{\Gamma}X$.\footnote{See Ref.~\cite{BoyleSmith:2026oay} for the role of the discrete torsion classified in terms of the bordism group of the Borel construction.}
Another direction is to study heterotic string theory.
As mentioned in Ref.~\cite{Yonekura:2022reu}, when the internal theory has a symmetry $G$, the relevant bordism groups are $\Omega^{\Spin}_{\ast}(BG\times B\O(10))$.

\section*{Acknowledgements}
The authors would like to thank Shuhei Ohyama, Ryohei Kobayashi, and Kazuya Yonekura for helpful discussions.
SK is especially grateful to Hayate Kimura for providing him with a mathematica code for computation of Steenrod squares.
The authors thank the RFCMP2026-S workshop (YITP-W-26-02) at the Yukawa Institute for Theoretical Physics, where this project was initiated.
The work of MK is supported by 
JSPS KAKENHI Grant Number 
25K23381, 25K01002, 
and 
the COREnet project (COREnet062) of
Research Center for Nuclear Physics, 
the University of Osaka.
The work of SS is supported by Forefront Physics and Mathematics Program to Drive Transformation (FoPM), a World-leading Innovative Graduate Study (WINGS) Program, the University of Tokyo, by Research Fellow of Japan Society for the Promotion of Science (JSPS Research Fellow), by JSPS KAKENHI Grant Number JP25KJ0857, and by JSR Fellowship, the University of Tokyo.
The work of HW is supported in part by JST FOREST Program (Grant Number JPMJFR2030, Japan).

\appendix

\section{Computation of the bordism groups}
\label{app:bordism}

In this appendix, we compute bordism groups using the Adams spectral sequence (ASS).
In many cases, computations of bordism groups are carried out using the Atiyah--Hirzebruch spectral sequence (AHSS) and the ASS together.
Compared with the AHSS, the ASS is more powerful and often determines even complicated extension problems.
In particular, all the bordism group computations needed in this paper, including the extension problems, can be carried out using the ASS.
For the concrete computational method of the ASS, see Refs.~\cite{Beaudry:2018ifm,Wan:2018bns,Wan:2019soo}.

Here we review some basic facts about the ASS.
The ASS was introduced by Adams in 1958 to compute the stable homotopy groups of topological spaces~\cite{Adams1958}.
Since generalized homology theories are represented by spectra and are obtained by computing stable homotopy groups of spectra, the ASS can also be applied to them.
In fact, the ASS converging to the generalized homology theory $E_*$ of a topological space $X'$ is given as follows:
\begin{equation}
    E_2^{s,t}=\mathrm{Ext}_{\cA_p}^{s,t}(\tilde{H}^*(E\wedge X';\bZ_p),\bZ_p)\implies\tilde{E}_{t-s}(X')_p^\wedge,
\end{equation}
where $\cA_p$ is the mod-$p$ Steenrod algebra, and ${}_p^\wedge$ means the $p$-completion.
In this paper, we consider the case for $p=2$ and $E\simeq M\Spin\wedge F$ where $F$ is $B\bZ_2$ or $\Sigma^1MTO(1)\wedge\Sigma^{-1}MO(1)$.
Then, using the K\"{u}nneth formula, the $E_2$ page is given by
\begin{equation}
    \mathrm{Ext}_{\cA_2}^{s,t}(\tilde{H}^*(M\Spin;\bZ_2)\otimes_{\bZ_2}\tilde{H}^*(F;\bZ_2)\otimes_{\bZ_2}\tilde{H}^*(X;\bZ_2),\bZ_2).
\end{equation}
In degrees at most seven, the cohomology of $M\Spin$ can be further simplified~\cite{ABP67,Freed:2016rqq}:
\begin{equation}
    \tilde{H}^*(M\Spin;\bZ_2)\cong\cA_2\otimes_{\cA_2(1)}(\bZ_2\oplus M_{\geq8}),
\end{equation}
where $\cA_2(1)$ is the subalgebra of $\cA_2$ generated by $\Sq^1$ and $\Sq^2$, and $M_{\ge8}$ is an $\cA_2(1)$-module with a degree greater than seven.
Therefore, using the adjunction formula, the $E_2$-page of the ASS in degrees less than eight is given as follows:
\begin{equation}
    \mathrm{Ext}_{\cA_2(1)}^{s,t}(\tilde{H}^*(F;\bZ_2)\otimes_{\bZ_2}\tilde{H}^*(X;\bZ_2),\bZ_2).
\end{equation}
Note that $\cA_2$ actions of the tensor product should be computed using the Cartan formula:
\begin{equation}
    \Sq^k(x\smile y) = \sum_{i+j=k} \Sq^i(x) \smile \Sq^j(y).
\end{equation}
In what follows, we use this formula to compute the bordism groups needed in this paper.
That is, in the discussion of worldsheet anomalies, we refer to the degree-three parts $\Omega^{\Spin\times\bZ_2}_3(B\O(10))$ and $\Omega^{\DPin}_3(B\O(10))$ for Type~II and Type~I, respectively.

\subsection{\texorpdfstring{$\Omega^{\Spin\times\bZ_2}_*(B\O(10))$}{Spin x Z2 bordism of BO(10)}}
\label{subsec:Adams_Type_II}

Here we compute the $\Spin\times\bZ_2$ bordism group for $B\O(10)$ using the following ASS:
\begin{equation}
    E_2^{s,t}=\mathrm{Ext}_{\cA_2(1)}^{s,t}(\tilde{H}^*(B\bZ_2\times B\O(10);\bZ_2),\bZ_2)\implies\tilde{\Omega}^{\Spin}_{t-s}(B\bZ_2\times B\O(10))_2^\wedge.
\end{equation}
Now the $\cA_2(1)$ module structure of $\tilde{H}^*(B\bZ_2\times B\O(10);\bZ_2)$ in low degrees is given by
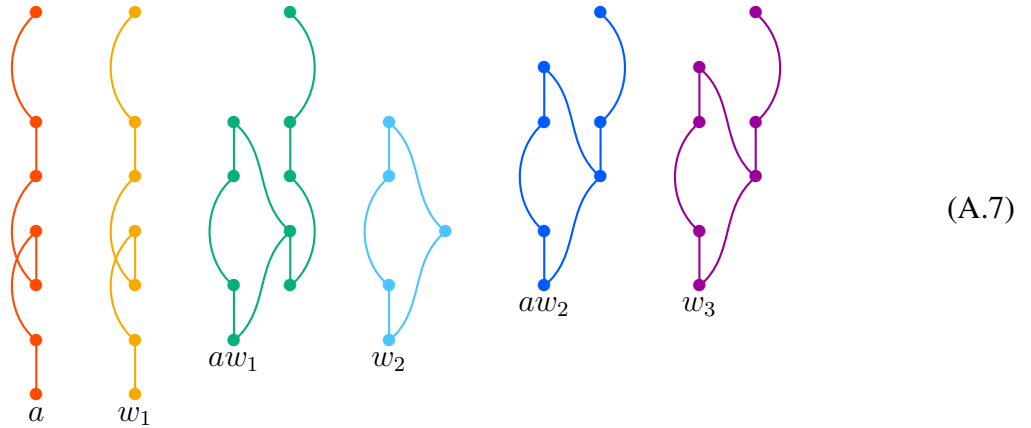
\begin{equation}
\vcenter{\hbox{
\begin{tikzpicture}[thick]
    \def\r{0.6em}
    \def\c{0.6em}
    \def\x{1.2em}
    
    \matrix (a) [matrix of math nodes,row sep=\r,column sep=\c,nodes=\colorA]{
        \bullet\\
        \phantom{\bullet}\\
        \bullet\\
        \bullet\\
        \bullet\\
        \bullet\\
        \bullet\\
        \bullet\\
    };
    \draw[\colorA] (a-1-1.center) to [out=220, in = 140] (a-3-1.center);
    \draw[\colorA] (a-4-1.center) to [out=220, in = 140] (a-6-1.center);
    \draw[\colorA] (a-5-1.center) to [out=220, in = 140] (a-7-1.center);
    \draw[\colorA] (a-3-1.center) to (a-4-1.center);
    \draw[\colorA] (a-5-1.center) to (a-6-1.center);
    \draw[\colorA] (a-7-1.center) to (a-8-1.center);
    \node[anchor=north] at (a-8-1) {$a$};
    
    \matrix (w1) [matrix of math nodes,row sep=\r,column sep=\c,nodes=\colorB,right=\x of a]{
        \bullet\\
        \phantom{\bullet}\\
        \bullet\\
        \bullet\\
        \bullet\\
        \bullet\\
        \bullet\\
        \bullet\\
    };
    \draw[\colorB] (w1-1-1.center) to [out=220, in = 140] (w1-3-1.center);
    \draw[\colorB] (w1-4-1.center) to [out=220, in = 140] (w1-6-1.center);
    \draw[\colorB] (w1-5-1.center) to [out=220, in = 140] (w1-7-1.center);
    \draw[\colorB] (w1-3-1.center) to (w1-4-1.center);
    \draw[\colorB] (w1-5-1.center) to (w1-6-1.center);
    \draw[\colorB] (w1-7-1.center) to (w1-8-1.center);
    \node[anchor=north] at (w1-8-1) {$w_1$};

    \matrix (aw1) [matrix of math nodes,row sep=\r,column sep=\c,nodes=\colorC,right=\x of w1]{
        \phantom{\bullet}&\bullet\\
        \phantom{\bullet}&\phantom{\bullet}\\
        \bullet&\bullet\\
        \bullet&\bullet\\
        \phantom{\bullet}&\bullet\\
        \bullet&\bullet\\
        \bullet&\phantom{\bullet}\\
        \phantom{\bullet}&\phantom{\bullet}\\
    };
    \draw[\colorC] (aw1-4-1.center) to [out=220, in = 140] (aw1-6-1.center);
    \draw[\colorC] (aw1-1-2.center) to [out=320, in =  40] (aw1-3-2.center);
    \draw[\colorC] (aw1-4-2.center) to [out=320, in =  40] (aw1-6-2.center);
    \draw[\colorC] (aw1-3-1.center) to [out=320, in = 140] (aw1-5-2.center);
    \draw[\colorC] (aw1-5-2.center) to [out=220, in =  40] (aw1-7-1.center);
    \draw[\colorC] (aw1-3-1.center) to (aw1-4-1.center);
    \draw[\colorC] (aw1-6-1.center) to (aw1-7-1.center);
    \draw[\colorC] (aw1-3-2.center) to (aw1-4-2.center);
    \draw[\colorC] (aw1-5-2.center) to (aw1-6-2.center);
    \node[anchor=north] at (aw1-7-1) {$aw_1$};

    \matrix (w2) [matrix of math nodes,row sep=\r,column sep=\c,nodes=\colorD,right=\x of aw1]{
        \phantom{\bullet}&\phantom{\bullet}\\
        \phantom{\bullet}&\phantom{\bullet}\\
        \bullet&\phantom{\bullet}\\
        \bullet&\phantom{\bullet}\\
        \phantom{\bullet}&\bullet\\
        \bullet&\phantom{\bullet}\\
        \bullet&\phantom{\bullet}\\
        \phantom{\bullet}&\phantom{\bullet}\\
    };
    \draw[\colorD] (w2-4-1.center) to [out=220, in = 140] (w2-6-1.center);
    \draw[\colorD] (w2-3-1.center) to [out=320, in = 140] (w2-5-2.center);
    \draw[\colorD] (w2-5-2.center) to [out=220, in =  40] (w2-7-1.center);
    \draw[\colorD] (w2-3-1.center) to (w2-4-1.center);
    \draw[\colorD] (w2-6-1.center) to (w2-7-1.center);
    \node[anchor=north] at (w2-7-1) {$w_2$};

    \matrix (aw2) [matrix of math nodes,row sep=\r,column sep=\c,nodes=\colorE,right=\x of w2]{
        \phantom{\bullet}&\bullet\\
        \bullet&\phantom{\bullet}\\
        \bullet&\bullet\\
        \phantom{\bullet}&\bullet\\
        \bullet&\phantom{\bullet}\\
        \bullet&\phantom{\bullet}\\
        \phantom{\bullet}&\phantom{\bullet}\\
        \phantom{\bullet}&\phantom{\bullet}\\
    };
    \draw[\colorE] (aw2-3-1.center) to [out=220, in = 140] (aw2-5-1.center);
    \draw[\colorE] (aw2-2-1.center) to [out=320, in = 140] (aw2-4-2.center);
    \draw[\colorE] (aw2-4-2.center) to [out=220, in =  40] (aw2-6-1.center);
    \draw[\colorE] (aw2-1-2.center) to [out=320, in =  40] (aw2-3-2.center);
    \draw[\colorE] (aw2-2-1.center) to (aw2-3-1.center);
    \draw[\colorE] (aw2-5-1.center) to (aw2-6-1.center);
    \draw[\colorE] (aw2-3-2.center) to (aw2-4-2.center);
    \node[anchor=north] at (aw2-6-1) {$aw_2$};

    \matrix (w3) [matrix of math nodes,row sep=\r,column sep=\c,nodes=\colorF,right=\x of aw2]{
        \phantom{\bullet}&\bullet\\
        \bullet&\phantom{\bullet}\\
        \bullet&\bullet\\
        \phantom{\bullet}&\bullet\\
        \bullet&\phantom{\bullet}\\
        \bullet&\phantom{\bullet}\\
        \phantom{\bullet}&\phantom{\bullet}\\
        \phantom{\bullet}&\phantom{\bullet}\\
    };
    \draw[\colorF] (w3-3-1.center) to [out=220, in = 140] (w3-5-1.center);
    \draw[\colorF] (w3-2-1.center) to [out=320, in = 140] (w3-4-2.center);
    \draw[\colorF] (w3-4-2.center) to [out=220, in =  40] (w3-6-1.center);
    \draw[\colorF] (w3-1-2.center) to [out=320, in =  40] (w3-3-2.center);
    \draw[\colorF] (w3-2-1.center) to (w3-3-1.center);
    \draw[\colorF] (w3-5-1.center) to (w3-6-1.center);
    \draw[\colorF] (w3-3-2.center) to (w3-4-2.center);
    \node[anchor=north] at (w3-6-1) {$w_3$};

\end{tikzpicture}
}}
\end{equation}
where the vertical lines represent the action of $\Sq^1$, and the curved lines represent the action of $\Sq^2$.
Also note that $a$ is the generator of $H^1(B\bZ_2;\bZ_2)$, and $w_i$ are the Stiefel--Whitney classes of $B\O(10)$.
Modules that are not connected to each other indicate a direct sum.
Thus, the Adams chart is given by
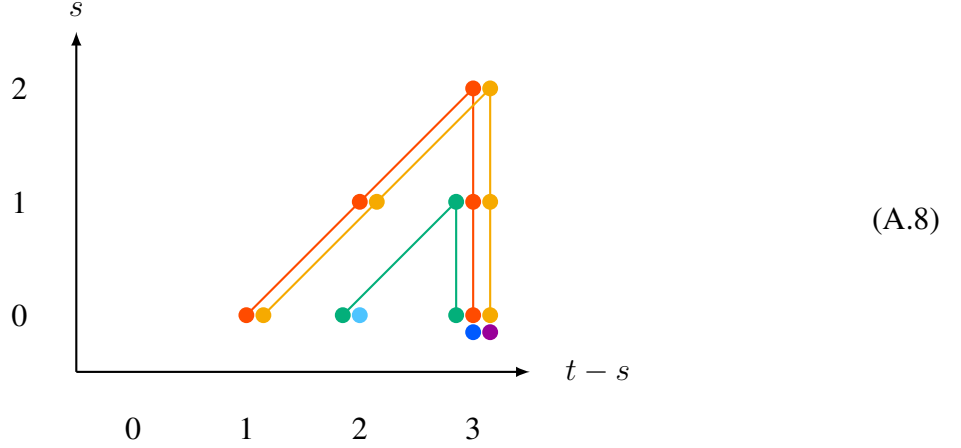
\begin{equation}
\vcenter{\hbox{
\begin{tikzpicture}[scale=1.5]
    \node at (0,-1) {0};
    \node at (1,-1) {1};
    \node at (2,-1) {2};
    \node at (3,-1) {3};
    \node at (4.1,-0.5) {$t-s$};
    \node at (-1,0) {0};
    \node at (-1,1) {1};
    \node at (-1,2) {2};
    \node at (-0.5,2.7) {$s$};
    \draw[-latex,thick] (-0.5,-0.5) -- (-0.5,2.5);
    \draw[-latex,thick] (-0.5,-0.5) -- (3.5,-0.5);

    \def\x{0.15}
    \def\s{0.7mm} 

    \fill[\colorA] (1,0) circle (\s);
    \fill[\colorA] (2,1) circle (\s);
    \fill[\colorA] (3,2) circle (\s);
    \fill[\colorA] (3,1) circle (\s);
    \fill[\colorA] (3,0) circle (\s);
    \draw[color=\colorA,thick] (1,0) -- (3,2);
    \draw[color=\colorA,thick] (3,2) -- (3,0);
    
    \fill[\colorB] (1+\x,0) circle (\s);
    \fill[\colorB] (2+\x,1) circle (\s);
    \fill[\colorB] (3+\x,2) circle (\s);
    \fill[\colorB] (3+\x,1) circle (\s);
    \fill[\colorB] (3+\x,0) circle (\s);
    \draw[color=\colorB,thick] (1+\x,0) -- (3+\x,2);
    \draw[color=\colorB,thick] (3+\x,2) -- (3+\x,0);
    
    \fill[\colorC] (2-\x,0) circle (\s);
    \fill[\colorC] (3-\x,1) circle (\s);
    \fill[\colorC] (3-\x,0) circle (\s);
    \draw[color=\colorC,thick] (2-\x,0) -- (3-\x,1);
    \draw[color=\colorC,thick] (3-\x,1) -- (3-\x,0);

    \fill[\colorD] (2,0) circle (\s);
    
    \fill[\colorE] (3,-\x) circle (\s);
    
    \fill[\colorF] (3+\x,-\x) circle (\s);
    
\end{tikzpicture}
}}
\end{equation}
where the vertical lines represent the action of $h_0\in\mathrm{Ext}_{\cA_2(1)}^{1,1}(\bZ_2,\bZ_2)$, while the diagonal lines represent the action of $h_1\in\mathrm{Ext}_{\cA_2(1)}^{1,2}(\bZ_2,\bZ_2)$.
The dots represent $\bZ_2$ factors, while the vertical lines represent nontrivial extensions between copies of $\bZ_2$.
That is, a line connecting two dots represents $\bZ_4$, while a line connecting three dots represents $\bZ_8$.
Finally, we find the following reduced bordism groups:
\begin{align}
    &\tilde{\Omega}^{\Spin\times\bZ_2}_0(B\O(10))=0 \\
    &\tilde{\Omega}^{\Spin\times\bZ_2}_1(B\O(10))=\bZ_2\langle w_1\rangle \\
    &\tilde{\Omega}^{\Spin\times\bZ_2}_2(B\O(10))=\bZ_2\langle aw_1\rangle\oplus \bZ_2\langle w_2\rangle\oplus \bZ_2\langle w_1^2\rangle \\
    &\tilde{\Omega}^{\Spin\times\bZ_2}_3(B\O(10))=\bZ_8\langle w_1^3\rangle\oplus \bZ_4\langle a^2w_1\rangle\oplus \bZ_2\langle aw_2\rangle\oplus \bZ_2\langle w_3\rangle
\end{align}
Note that, by an abuse of notation, we use the same symbol for the corresponding generator in cohomology.
We use the relation $\tilde{\Omega}^{\Spin\times\bZ_2}_*(B\O(10))\cong\tilde{\Omega}^{\Spin}_*(B\O(10))\oplus\tilde{\Omega}^{\Spin}_*(B\bZ_2\wedge B\O(10))$ and remove the contribution from $\tilde{\Omega}^{\Spin}_*(B\bZ_2)$, which appears in $\tilde{\Omega}^{\Spin}_*(B\bZ_2\times B\O(10))$.
It is clear that the bordism groups are trivial at all other odd primes.

\subsection{\texorpdfstring{$\Omega^{\mathrm{DPin}}_*(B\O(10))$}{DPin bordism of BO(10)}}
\label{subsec:Adams_Type_I}

In this subsection, we compute the bordism groups $\Omega^{\DPin}_{d}(B\O(10))$ up to $d=3$ using the ASS.
According to Ref.~\cite{Wan:2019soo}, the spectrum $MT\mathrm{DPin}$ decomposes as follows:
\begin{equation}
    MT\mathrm{DPin}\simeq M\Spin\wedge\Sigma^1MT\O(1)\wedge\Sigma^{-1}M\O(1).
\end{equation}
Note that $MT\Pin^+\simeq M\Spin\wedge\Sigma^1MT\O(1)$ and $MT\Pin^-\simeq M\Spin\wedge\Sigma^{-1}M\O(1)$, and our computation is directly applicable to the case of unoriented Type~0 string theories.
Then, the $\mathrm{DPin}$ bordism groups for $B\O(10)$ can be computed using the following ASS:
\begin{equation}
    E_2^{s,t} = \text{Ext}^{s,t}_{\mathcal{A}_2(1)} (M,\mathbb{Z}_2)\Rightarrow \tilde{\Omega}^{\text{DPin}}_{t-s}(B\O(10))_2^\wedge.
\end{equation}
where $M$ is the following $\cA_2(1)$ module
\begin{equation}
    M=\tilde{H}^{*-1}(MT\O(1);\mathbb{Z}_2) \otimes \tilde{H}^{*+1}(M\O(1);\mathbb{Z}_2) \otimes \tilde{H}^*(B\O(10);\mathbb{Z}_2)
\end{equation}
The $\cA_2(1)$ module structure of $M$ in the range we focus on is given by
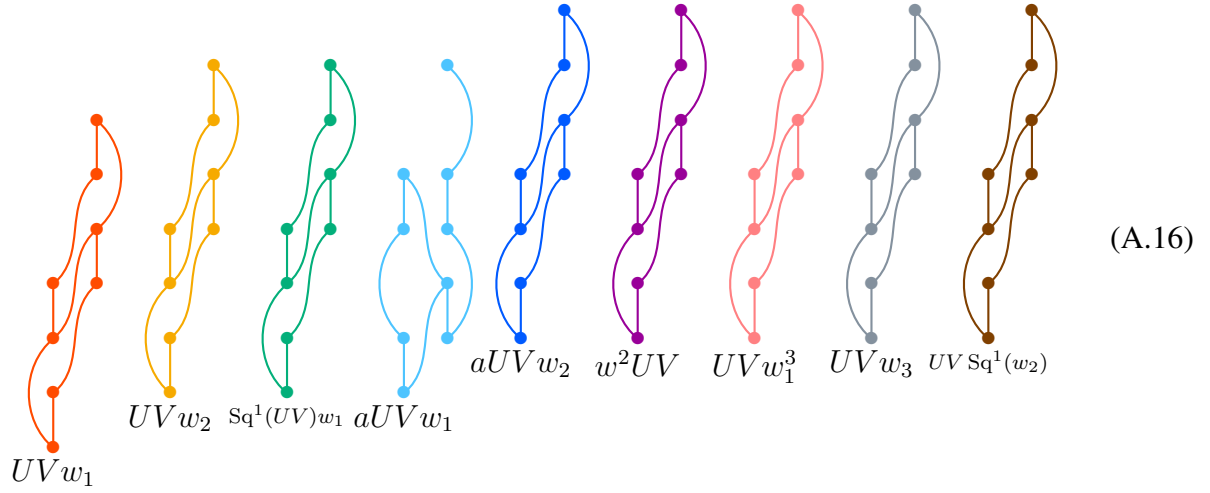
\begin{equation}
\vcenter{\hbox{
\begin{tikzpicture}[thick]
    \def\r{0.6em}
    \def\c{0.2em}
    \def\x{0.4em}
    
    \matrix (UVw1) [matrix of math nodes,row sep=\r,column sep=\c,nodes=\colorA]{
        \phantom{\bullet}&\phantom{\bullet}\\
        \phantom{\bullet}&\phantom{\bullet}\\
        \phantom{\bullet}&\bullet\\
        \phantom{\bullet}&\bullet\\
        \phantom{\bullet}&\bullet\\
        \bullet&\bullet\\
        \bullet&\phantom{\bullet}\\
        \bullet&\phantom{\bullet}\\
        \bullet&\phantom{\bullet}\\
    };
    \draw[\colorA] (UVw1-7-1.center) to [out=220, in = 140] (UVw1-9-1.center);
    \draw[\colorA] (UVw1-3-2.center) to [out=320, in =  40] (UVw1-5-2.center);
    \draw[\colorA] (UVw1-4-2.center) to [out=220, in =  40] (UVw1-6-1.center);
    \draw[\colorA] (UVw1-5-2.center) to [out=220, in =  40] (UVw1-7-1.center);
    \draw[\colorA] (UVw1-6-2.center) to [out=220, in =  40] (UVw1-8-1.center);
    \draw[\colorA] (UVw1-6-1.center) to (UVw1-7-1.center);
    \draw[\colorA] (UVw1-8-1.center) to (UVw1-9-1.center);
    \draw[\colorA] (UVw1-3-2.center) to (UVw1-4-2.center);
    \draw[\colorA] (UVw1-5-2.center) to (UVw1-6-2.center);
    \node[anchor=north] at (UVw1-9-1) {$UVw_1$};
    
    \matrix (UVw2) [matrix of math nodes,row sep=\r,column sep=\c,nodes=\colorB,right=\x of UVw1]{
        \phantom{\bullet}&\phantom{\bullet}\\
        \phantom{\bullet}&\bullet\\
        \phantom{\bullet}&\bullet\\
        \phantom{\bullet}&\bullet\\
        \bullet&\bullet\\
        \bullet&\phantom{\bullet}\\
        \bullet&\phantom{\bullet}\\
        \bullet&\phantom{\bullet}\\
        \phantom{\bullet}&\phantom{\bullet}\\
    };
    \draw[\colorB] (UVw2-6-1.center) to [out=220, in = 140] (UVw2-8-1.center);
    \draw[\colorB] (UVw2-2-2.center) to [out=320, in =  40] (UVw2-4-2.center);
    \draw[\colorB] (UVw2-3-2.center) to [out=220, in =  40] (UVw2-5-1.center);
    \draw[\colorB] (UVw2-4-2.center) to [out=220, in =  40] (UVw2-6-1.center);
    \draw[\colorB] (UVw2-5-2.center) to [out=220, in =  40] (UVw2-7-1.center);
    \draw[\colorB] (UVw2-5-1.center) to (UVw2-6-1.center);
    \draw[\colorB] (UVw2-7-1.center) to (UVw2-8-1.center);
    \draw[\colorB] (UVw2-2-2.center) to (UVw2-3-2.center);
    \draw[\colorB] (UVw2-4-2.center) to (UVw2-5-2.center);
    \node[anchor=north] at (UVw2-8-1) {$UVw_2$};
    
    \matrix (Sq1UVw1) [matrix of math nodes,row sep=\r,column sep=\c,nodes=\colorC,right=\x of UVw2]{
        \phantom{\bullet}&\phantom{\bullet}\\
        \phantom{\bullet}&\bullet\\
        \phantom{\bullet}&\bullet\\
        \phantom{\bullet}&\bullet\\
        \bullet&\bullet\\
        \bullet&\phantom{\bullet}\\
        \bullet&\phantom{\bullet}\\
        \bullet&\phantom{\bullet}\\
        \phantom{\bullet}&\phantom{\bullet}\\
    };
    \draw[\colorC] (Sq1UVw1-6-1.center) to [out=220, in = 140] (Sq1UVw1-8-1.center);
    \draw[\colorC] (Sq1UVw1-2-2.center) to [out=320, in =  40] (Sq1UVw1-4-2.center);
    \draw[\colorC] (Sq1UVw1-3-2.center) to [out=220, in =  40] (Sq1UVw1-5-1.center);
    \draw[\colorC] (Sq1UVw1-4-2.center) to [out=220, in =  40] (Sq1UVw1-6-1.center);
    \draw[\colorC] (Sq1UVw1-5-2.center) to [out=220, in =  40] (Sq1UVw1-7-1.center);
    \draw[\colorC] (Sq1UVw1-5-1.center) to (Sq1UVw1-6-1.center);
    \draw[\colorC] (Sq1UVw1-7-1.center) to (Sq1UVw1-8-1.center);
    \draw[\colorC] (Sq1UVw1-2-2.center) to (Sq1UVw1-3-2.center);
    \draw[\colorC] (Sq1UVw1-4-2.center) to (Sq1UVw1-5-2.center);
    \node[anchor=north] at (Sq1UVw1-8-1) {\scriptsize $\Sq^1(UV)w_1$};

    \matrix (aUVw1) [matrix of math nodes,row sep=\r,column sep=\c,nodes=\colorD,right=\x of Sq1UVw1]{
        \phantom{\bullet}&\phantom{\bullet}\\
        \phantom{\bullet}&\bullet\\
        \phantom{\bullet}&\phantom{\bullet}\\
        \bullet&\bullet\\
        \bullet&\bullet\\
        \phantom{\bullet}&\bullet\\
        \bullet&\bullet\\
        \bullet&\phantom{\bullet}\\
        \phantom{\bullet}&\phantom{\bullet}\\
    };
    \draw[\colorD] (aUVw1-5-1.center) to [out=220, in = 140] (aUVw1-7-1.center);
    \draw[\colorD] (aUVw1-2-2.center) to [out=320, in =  40] (aUVw1-4-2.center);
    \draw[\colorD] (aUVw1-5-2.center) to [out=320, in =  40] (aUVw1-7-2.center);
    \draw[\colorD] (aUVw1-4-1.center) to [out=320, in = 140] (aUVw1-6-2.center);
    \draw[\colorD] (aUVw1-6-2.center) to [out=220, in =  40] (aUVw1-8-1.center);
    \draw[\colorD] (aUVw1-4-1.center) to (aUVw1-5-1.center);
    \draw[\colorD] (aUVw1-7-1.center) to (aUVw1-8-1.center);
    \draw[\colorD] (aUVw1-4-2.center) to (aUVw1-5-2.center);
    \draw[\colorD] (aUVw1-6-2.center) to (aUVw1-7-2.center);
    \node[anchor=north] at (aUVw1-8-1) {$aUVw_1$};

    \matrix (aUVw2) [matrix of math nodes,row sep=\r,column sep=\c,nodes=\colorE,right=\x of aUVw1]{
        \phantom{\bullet}&\bullet\\
        \phantom{\bullet}&\bullet\\
        \phantom{\bullet}&\bullet\\
        \bullet&\bullet\\
        \bullet&\phantom{\bullet}\\
        \bullet&\phantom{\bullet}\\
        \bullet&\phantom{\bullet}\\
        \phantom{\bullet}&\phantom{\bullet}\\
        \phantom{\bullet}&\phantom{\bullet}\\
    };
    \draw[\colorE] (aUVw2-5-1.center) to [out=220, in = 140] (aUVw2-7-1.center);
    \draw[\colorE] (aUVw2-1-2.center) to [out=320, in =  40] (aUVw2-3-2.center);
    \draw[\colorE] (aUVw2-2-2.center) to [out=220, in =  40] (aUVw2-4-1.center);
    \draw[\colorE] (aUVw2-3-2.center) to [out=220, in =  40] (aUVw2-5-1.center);
    \draw[\colorE] (aUVw2-4-2.center) to [out=220, in =  40] (aUVw2-6-1.center);
    \draw[\colorE] (aUVw2-4-1.center) to (aUVw2-5-1.center);
    \draw[\colorE] (aUVw2-6-1.center) to (aUVw2-7-1.center);
    \draw[\colorE] (aUVw2-1-2.center) to (aUVw2-2-2.center);
    \draw[\colorE] (aUVw2-3-2.center) to (aUVw2-4-2.center);
    \node[anchor=north] at (aUVw2-7-1) {$aUVw_2$};

    \matrix (w2UV) [matrix of math nodes,row sep=\r,column sep=\c,nodes=\colorF,right=\x of aUVw2]{
        \phantom{\bullet}&\bullet\\
        \phantom{\bullet}&\bullet\\
        \phantom{\bullet}&\bullet\\
        \bullet&\bullet\\
        \bullet&\phantom{\bullet}\\
        \bullet&\phantom{\bullet}\\
        \bullet&\phantom{\bullet}\\
        \phantom{\bullet}&\phantom{\bullet}\\
        \phantom{\bullet}&\phantom{\bullet}\\
    };
    \draw[\colorF] (w2UV-5-1.center) to [out=220, in = 140] (w2UV-7-1.center);
    \draw[\colorF] (w2UV-1-2.center) to [out=320, in =  40] (w2UV-3-2.center);
    \draw[\colorF] (w2UV-2-2.center) to [out=220, in =  40] (w2UV-4-1.center);
    \draw[\colorF] (w2UV-3-2.center) to [out=220, in =  40] (w2UV-5-1.center);
    \draw[\colorF] (w2UV-4-2.center) to [out=220, in =  40] (w2UV-6-1.center);
    \draw[\colorF] (w2UV-4-1.center) to (w2UV-5-1.center);
    \draw[\colorF] (w2UV-6-1.center) to (w2UV-7-1.center);
    \draw[\colorF] (w2UV-1-2.center) to (w2UV-2-2.center);
    \draw[\colorF] (w2UV-3-2.center) to (w2UV-4-2.center);
    \node[anchor=north] at (w2UV-7-1) {$w^2UV$};

    \matrix (UVw13) [matrix of math nodes,row sep=\r,column sep=\c,nodes=\colorG,right=\x of w2UV]{
        \phantom{\bullet}&\bullet\\
        \phantom{\bullet}&\bullet\\
        \phantom{\bullet}&\bullet\\
        \bullet&\bullet\\
        \bullet&\phantom{\bullet}\\
        \bullet&\phantom{\bullet}\\
        \bullet&\phantom{\bullet}\\
        \phantom{\bullet}&\phantom{\bullet}\\
        \phantom{\bullet}&\phantom{\bullet}\\
    };
    \draw[\colorG] (UVw13-5-1.center) to [out=220, in = 140] (UVw13-7-1.center);
    \draw[\colorG] (UVw13-1-2.center) to [out=320, in =  40] (UVw13-3-2.center);
    \draw[\colorG] (UVw13-2-2.center) to [out=220, in =  40] (UVw13-4-1.center);
    \draw[\colorG] (UVw13-3-2.center) to [out=220, in =  40] (UVw13-5-1.center);
    \draw[\colorG] (UVw13-4-2.center) to [out=220, in =  40] (UVw13-6-1.center);
    \draw[\colorG] (UVw13-4-1.center) to (UVw13-5-1.center);
    \draw[\colorG] (UVw13-6-1.center) to (UVw13-7-1.center);
    \draw[\colorG] (UVw13-1-2.center) to (UVw13-2-2.center);
    \draw[\colorG] (UVw13-3-2.center) to (UVw13-4-2.center);
    \node[anchor=north] at (UVw13-7-1) {$UVw_1^3$};

    \matrix (UVw3) [matrix of math nodes,row sep=\r,column sep=\c,nodes=\colorH,right=\x of UVw13]{
        \phantom{\bullet}&\bullet\\
        \phantom{\bullet}&\bullet\\
        \phantom{\bullet}&\bullet\\
        \bullet&\bullet\\
        \bullet&\phantom{\bullet}\\
        \bullet&\phantom{\bullet}\\
        \bullet&\phantom{\bullet}\\
        \phantom{\bullet}&\phantom{\bullet}\\
        \phantom{\bullet}&\phantom{\bullet}\\
    };
    \draw[\colorH] (UVw3-5-1.center) to [out=220, in = 140] (UVw3-7-1.center);
    \draw[\colorH] (UVw3-1-2.center) to [out=320, in =  40] (UVw3-3-2.center);
    \draw[\colorH] (UVw3-2-2.center) to [out=220, in =  40] (UVw3-4-1.center);
    \draw[\colorH] (UVw3-3-2.center) to [out=220, in =  40] (UVw3-5-1.center);
    \draw[\colorH] (UVw3-4-2.center) to [out=220, in =  40] (UVw3-6-1.center);
    \draw[\colorH] (UVw3-4-1.center) to (UVw3-5-1.center);
    \draw[\colorH] (UVw3-6-1.center) to (UVw3-7-1.center);
    \draw[\colorH] (UVw3-1-2.center) to (UVw3-2-2.center);
    \draw[\colorH] (UVw3-3-2.center) to (UVw3-4-2.center);
    \node[anchor=north] at (UVw3-7-1) {$UVw_3$};

    \matrix (UVSq1w2) [matrix of math nodes,row sep=\r,column sep=\c,nodes=\colorI,right=\x of UVw3]{
        \phantom{\bullet}&\bullet\\
        \phantom{\bullet}&\bullet\\
        \phantom{\bullet}&\bullet\\
        \bullet&\bullet\\
        \bullet&\phantom{\bullet}\\
        \bullet&\phantom{\bullet}\\
        \bullet&\phantom{\bullet}\\
        \phantom{\bullet}&\phantom{\bullet}\\
        \phantom{\bullet}&\phantom{\bullet}\\
    };
    \draw[\colorI] (UVSq1w2-5-1.center) to [out=220, in = 140] (UVSq1w2-7-1.center);
    \draw[\colorI] (UVSq1w2-1-2.center) to [out=320, in =  40] (UVSq1w2-3-2.center);
    \draw[\colorI] (UVSq1w2-2-2.center) to [out=220, in =  40] (UVSq1w2-4-1.center);
    \draw[\colorI] (UVSq1w2-3-2.center) to [out=220, in =  40] (UVSq1w2-5-1.center);
    \draw[\colorI] (UVSq1w2-4-2.center) to [out=220, in =  40] (UVSq1w2-6-1.center);
    \draw[\colorI] (UVSq1w2-4-1.center) to (UVSq1w2-5-1.center);
    \draw[\colorI] (UVSq1w2-6-1.center) to (UVSq1w2-7-1.center);
    \draw[\colorI] (UVSq1w2-1-2.center) to (UVSq1w2-2-2.center);
    \draw[\colorI] (UVSq1w2-3-2.center) to (UVSq1w2-4-2.center);
    \node[anchor=north] at (UVSq1w2-7-1) {\scriptsize $UV\Sq^1(w_2)$};

\end{tikzpicture}
}}
\end{equation}
Again, we use the notation where the vertical lines represent the action of $\Sq^1$, and the curved lines represent the action of $\Sq^2$.
Note that $H^{*-1}(MT\O(1);\mathbb{Z}_2) \cong \bZ_2[a] U$ and $H^{*+1} (M\O(1),\mathbb{Z}_2) \cong \mathbb{Z}_2 [w] V$.
The minimal resolution of the module isomorphic to $\cA_2(1)$ terminates immediately, yielding a single point in the Adams chart.
Therefore, the entire Adams chart is given by
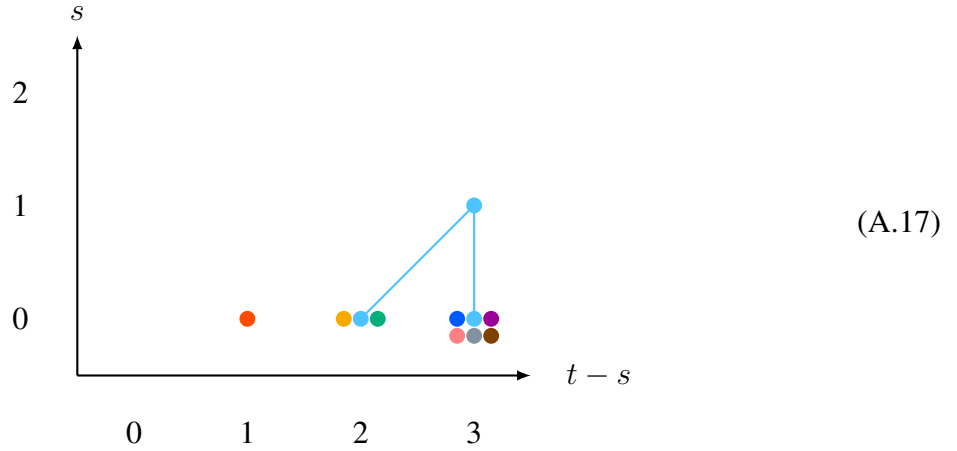
\begin{equation}
\vcenter{\hbox{
\begin{tikzpicture}[scale=1.5]
    \node at (0,-1) {0};
    \node at (1,-1) {1};
    \node at (2,-1) {2};
    \node at (3,-1) {3};
    \node at (4.1,-0.5) {$t-s$};
    \node at (-1,0) {0};
    \node at (-1,1) {1};
    \node at (-1,2) {2};
    \node at (-0.5,2.7) {$s$};
    \draw[-latex,thick] (-0.5,-0.5) -- (-0.5,2.5);
    \draw[-latex,thick] (-0.5,-0.5) -- (3.5,-0.5);

    \def\x{0.15}
    \def\s{0.7mm} 
    
    \fill[\colorA] (1,0) circle (\s);
    
    \fill[\colorB] (2-\x,0) circle (\s);
    
    \fill[\colorC] (2+\x,0) circle (\s);

    \fill[\colorD] (2,0) circle (\s);
    \fill[\colorD] (3,1) circle (\s);
    \fill[\colorD] (3,0) circle (\s);
    \draw[color=\colorD,thick] (2,0) -- (3,1);
    \draw[color=\colorD,thick] (3,1) -- (3,0);
    
    \fill[\colorE] (3-\x,0) circle (\s);
    
    \fill[\colorF] (3+\x,0) circle (\s);
    
    \fill[\colorG] (3-\x,-\x) circle (\s);
    
    \fill[\colorH] (3,-\x) circle (\s);
    
    \fill[\colorI] (3+\x,-\x) circle (\s);
    
\end{tikzpicture}
}}
\end{equation}
where the vertical lines represent the action of $h_0$, while the diagonal lines represent the action of $h_1$.
Therefore, the reduced bordism groups are given by
\begin{align}
    &\tilde{\Omega}^{\mathrm{DPin}}_0(B\O(10))=0 \\
    &\tilde{\Omega}^{\mathrm{DPin}}_1(B\O(10))=\bZ_2\langle UVw_1\rangle \\
    &\tilde{\Omega}^{\mathrm{DPin}}_2(B\O(10))=\bZ_2\langle UVw_2\rangle\oplus  \bZ_2\langle aUVw_1\rangle\oplus  \bZ_2\langle UVw_1^2\rangle \\
    &\tilde{\Omega}^{\mathrm{DPin}}_3(B\O(10))=\bZ_4\langle aUVw_1^2\rangle\oplus \bZ_2\langle aUVw_2\rangle\oplus \bZ_2\langle w^2UVw_1\rangle \notag \\
    &\qquad\qquad\qquad\qquad\oplus\bZ_2\langle UVw_1^3\rangle\oplus  \bZ_2\langle UVw_3\rangle\oplus  \bZ_2\langle UVw_1w_2\rangle
\end{align}
Again, we use the same symbol for the corresponding generator in cohomology.

\newpage

\bibliographystyle{ytphys}
\bibliography{main}

\providecommand{\href}[2]{#2}\begingroup\raggedright\begin{thebibliography}{10}

\bibitem{Neveu:1971rx}
A.~Neveu and J.~H. Schwarz, {\slshape {Factorizable dual model of pions},} \href{http://dx.doi.org/10.1016/0550-3213(71)90448-2}{{\em Nucl. Phys. B} {\bfseries 31} (1971) 86--112}.

\bibitem{Ramond:1971gb}
P.~Ramond, {\slshape {Dual Theory for Free Fermions},} \href{http://dx.doi.org/10.1103/PhysRevD.3.2415}{{\em Phys. Rev. D} {\bfseries 3} (1971) 2415--2418}.

\bibitem{Gliozzi:1976jf}
F.~Gliozzi, J.~Scherk, and D.~I. Olive, {\slshape {Supergravity and the Spinor Dual Model},} \href{http://dx.doi.org/10.1016/0370-2693(76)90183-0}{{\em Phys. Lett. B} {\bfseries 65} (1976) 282--286}.

\bibitem{Gliozzi:1976qd}
F.~Gliozzi, J.~Scherk, and D.~I. Olive, {\slshape {Supersymmetry, Supergravity Theories and the Dual Spinor Model},} \href{http://dx.doi.org/10.1016/0550-3213(77)90206-1}{{\em Nucl. Phys. B} {\bfseries 122} (1977) 253--290}.

\bibitem{Moore:1984dc}
G.~W. Moore and P.~C. Nelson, {\slshape {Anomalies in Nonlinear $\sigma$ Models},} \href{http://dx.doi.org/10.1103/PhysRevLett.53.1519}{{\em Phys. Rev. Lett.} {\bfseries 53} (1984) 1519}.

\bibitem{Moore:1984ws}
G.~W. Moore and P.~C. Nelson, {\slshape {The Etiology of $\sigma$ Model Anomalies},} \href{http://dx.doi.org/10.1007/BF01212688}{{\em Commun. Math. Phys.} {\bfseries 100} (1985) 83}.

\bibitem{Freed:2004yc}
D.~S. Freed and G.~W. Moore, {\slshape {Setting the quantum integrand of M-theory},} \href{http://dx.doi.org/10.1007/s00220-005-1482-7}{{\em Commun. Math. Phys.} {\bfseries 263} (2006) 89--132}, \href{http://arxiv.org/abs/hep-th/0409135}{{ arXiv:hep-th/0409135}}.

\bibitem{Callan:1984sa}
C.~G. Callan, Jr. and J.~A. Harvey, {\slshape {Anomalies and Fermion Zero Modes on Strings and Domain Walls},} \href{http://dx.doi.org/10.1016/0550-3213(85)90489-4}{{\em Nucl. Phys. B} {\bfseries 250} (1985) 427--436}.

\bibitem{Aminov:2026zrv}
G.~Aminov, C.~Cs{\'a}ki, O.~Telem, and S.~Yankielowicz, {\slshape {Non-Abelian and type-A conformal anomalies from Euler descent},} \href{http://dx.doi.org/10.1007/JHEP07(2026)053}{{\em JHEP} {\bfseries 07} (2026) 053}, \href{http://arxiv.org/abs/2601.18892}{{ arXiv:2601.18892~[hep-th]}}.

\bibitem{Witten:2015aba}
E.~Witten, {\slshape {Fermion Path Integrals And Topological Phases},} \href{http://dx.doi.org/10.1103/RevModPhys.88.035001}{{\em Rev. Mod. Phys.} {\bfseries 88} (2016) 035001}, \href{http://arxiv.org/abs/1508.04715}{{ arXiv:1508.04715~[cond-mat.mes-hall]}}.

\bibitem{Witten:2019bou}
E.~Witten and K.~Yonekura, {\slshape {Anomaly Inflow and the $\eta$-Invariant},} in {\em {The Shoucheng Zhang Memorial Workshop}}.
\newblock 9, 2019.
\newblock \href{http://arxiv.org/abs/1909.08775}{{ arXiv:1909.08775~[hep-th]}}.

\bibitem{Kapustin:2014tfa}
A.~Kapustin, {\slshape {Symmetry Protected Topological Phases, Anomalies, and Cobordisms: Beyond Group Cohomology},} \href{http://arxiv.org/abs/1403.1467}{{ arXiv:1403.1467~[cond-mat.str-el]}}.

\bibitem{Kapustin:2014dxa}
A.~Kapustin, R.~Thorngren, A.~Turzillo, and Z.~Wang, {\slshape {Fermionic Symmetry Protected Topological Phases and Cobordisms},} \href{http://dx.doi.org/10.1007/JHEP12(2015)052}{{\em JHEP} {\bfseries 12} (2015) 052}, \href{http://arxiv.org/abs/1406.7329}{{ arXiv:1406.7329~[cond-mat.str-el]}}.

\bibitem{Freed:2016rqq}
D.~S. Freed and M.~J. Hopkins, {\slshape {Reflection positivity and invertible topological phases},} \href{http://dx.doi.org/10.2140/gt.2021.25.1165}{{\em Geom. Topol.} {\bfseries 25} (2021) 1165--1330}, \href{http://arxiv.org/abs/1604.06527}{{ arXiv:1604.06527~[hep-th]}}.

\bibitem{Yonekura:2018ufj}
K.~Yonekura, {\slshape {On the cobordism classification of symmetry protected topological phases},} \href{http://dx.doi.org/10.1007/s00220-019-03439-y}{{\em Commun. Math. Phys.} {\bfseries 368} (2019) 1121--1173}, \href{http://arxiv.org/abs/1803.10796}{{ arXiv:1803.10796~[hep-th]}}.

\bibitem{Sagnotti:1987tw}
A.~Sagnotti, {\slshape {Open Strings and their Symmetry Groups},} in {\em {NATO Advanced Summer Institute on Nonperturbative Quantum Field Theory (Cargese Summer Institute)}}.
\newblock 9, 1987.
\newblock \href{http://arxiv.org/abs/hep-th/0208020}{{ arXiv:hep-th/0208020}}.

\bibitem{Dai:1989ua}
J.~Dai, R.~G. Leigh, and J.~Polchinski, {\slshape {New Connections Between String Theories},} \href{http://dx.doi.org/10.1142/S0217732389002331}{{\em Mod. Phys. Lett. A} {\bfseries 4} (1989) 2073--2083}.

\bibitem{Vafa:1986wx}
C.~Vafa, {\slshape {Modular Invariance and Discrete Torsion on Orbifolds},} \href{http://dx.doi.org/10.1016/0550-3213(86)90379-2}{{\em Nucl. Phys. B} {\bfseries 273} (1986) 592--606}.

\bibitem{Freed:1999vc}
D.~S. Freed and E.~Witten, {\slshape {Anomalies in string theory with D-branes},} {\em Asian J. Math.} {\bfseries 3} (1999) 819, \href{http://arxiv.org/abs/hep-th/9907189}{{ arXiv:hep-th/9907189}}.

\bibitem{Witten:1999eg}
E.~Witten, {\slshape {World-sheet corrections via $D$-instantons.},} \href{http://dx.doi.org/10.1088/1126-6708/2000/02/030}{{\em JHEP} {\bfseries 02} (2000) 030}, \href{http://arxiv.org/abs/hep-th/9907041}{{ arXiv:hep-th/9907041}}.

\bibitem{Dixon:1986iz}
L.~J. Dixon and J.~A. Harvey, {\slshape {String Theories in Ten-Dimensions Without Space-Time Supersymmetry},} \href{http://dx.doi.org/10.1016/0550-3213(86)90619-X}{{\em Nucl. Phys. B} {\bfseries 274} (1986) 93--105}.

\bibitem{Bianchi:1990yu}
M.~Bianchi and A.~Sagnotti, {\slshape {On the systematics of open string theories},} \href{http://dx.doi.org/10.1016/0370-2693(90)91894-H}{{\em Phys. Lett. B} {\bfseries 247} (1990) 517--524}.

\bibitem{Sagnotti:1996qj}
A.~Sagnotti, {\slshape {Surprises in open string perturbation theory},} \href{http://dx.doi.org/10.1016/S0920-5632(97)00344-7}{{\em Nucl. Phys. B Proc. Suppl.} {\bfseries 56} (1997) 332--343}, \href{http://arxiv.org/abs/hep-th/9702093}{{ arXiv:hep-th/9702093}}.

\bibitem{Bergman:1997rf}
O.~Bergman and M.~R. Gaberdiel, {\slshape {A Nonsupersymmetric open string theory and S duality},} \href{http://dx.doi.org/10.1016/S0550-3213(97)00309-X}{{\em Nucl. Phys. B} {\bfseries 499} (1997) 183--204}, \href{http://arxiv.org/abs/hep-th/9701137}{{ arXiv:hep-th/9701137}}.

\bibitem{Bergman:1999km}
O.~Bergman and M.~R. Gaberdiel, {\slshape {Dualities of type 0 strings},} \href{http://dx.doi.org/10.1088/1126-6708/1999/07/022}{{\em JHEP} {\bfseries 07} (1999) 022}, \href{http://arxiv.org/abs/hep-th/9906055}{{ arXiv:hep-th/9906055}}.

\bibitem{Blumenhagen:1999ad}
R.~Blumenhagen and A.~Kumar, {\slshape {A Note on orientifolds and dualities of type 0B string theory},} \href{http://dx.doi.org/10.1016/S0370-2693(99)01002-3}{{\em Phys. Lett. B} {\bfseries 464} (1999) 46--52}, \href{http://arxiv.org/abs/hep-th/9906234}{{ arXiv:hep-th/9906234}}.

\bibitem{Kaidi:2019pzj}
J.~Kaidi, J.~Parra-Martinez, and Y.~Tachikawa, {\slshape {Classification of String Theories via Topological Phases},} \href{http://dx.doi.org/10.1103/PhysRevLett.124.121601}{{\em Phys. Rev. Lett.} {\bfseries 124} (2020) 121601}, \href{http://arxiv.org/abs/1908.04805}{{ arXiv:1908.04805~[hep-th]}}.

\bibitem{Kaidi:2019tyf}
J.~Kaidi, J.~Parra-Martinez, and Y.~Tachikawa, {\slshape {Topological Superconductors on Superstring Worldsheets},} \href{http://dx.doi.org/10.21468/SciPostPhys.9.1.010}{{\em SciPost Phys.} {\bfseries 9} (2020) 10}, \href{http://arxiv.org/abs/1911.11780}{{ arXiv:1911.11780~[hep-th]}}.

\bibitem{Witten:2023snr}
E.~Witten, {\slshape {Anomalies and Nonsupersymmetric D-Branes},} \href{http://arxiv.org/abs/2305.01012}{{ arXiv:2305.01012~[hep-th]}}.

\bibitem{Delgado:2026qvy}
M.~Delgado, L.~Eberhardt, and M.~Toma{\v{s}}evi{\'c}, {\slshape {What makes spacetime spin in string theory?},} \href{http://arxiv.org/abs/2606.18380}{{ arXiv:2606.18380~[hep-th]}}.

\bibitem{Yonekura:2022reu}
K.~Yonekura, {\slshape {Heterotic global anomalies and torsion Witten index},} \href{http://dx.doi.org/10.1007/JHEP10(2022)114}{{\em JHEP} {\bfseries 10} (2022) 114}, \href{http://arxiv.org/abs/2207.13858}{{ arXiv:2207.13858~[hep-th]}}.

\bibitem{Gross:1984dd}
D.~J. Gross, J.~A. Harvey, E.~J. Martinec, and R.~Rohm, {\slshape {The Heterotic String},} \href{http://dx.doi.org/10.1103/PhysRevLett.54.502}{{\em Phys. Rev. Lett.} {\bfseries 54} (1985) 502--505}.

\bibitem{Gross:1985fr}
D.~J. Gross, J.~A. Harvey, E.~J. Martinec, and R.~Rohm, {\slshape {Heterotic String Theory. 1. The Free Heterotic String},} \href{http://dx.doi.org/10.1016/0550-3213(85)90394-3}{{\em Nucl. Phys. B} {\bfseries 256} (1985) 253}.

\bibitem{Gross:1985rr}
D.~J. Gross, J.~A. Harvey, E.~J. Martinec, and R.~Rohm, {\slshape {Heterotic String Theory. 2. The Interacting Heterotic String},} \href{http://dx.doi.org/10.1016/0550-3213(86)90146-X}{{\em Nucl. Phys. B} {\bfseries 267} (1986) 75--124}.

\bibitem{Yamashita:2021cao}
M.~Yamashita and K.~Yonekura, {\slshape {Differential models for the Anderson dual to bordism theories and invertible QFT{\textquoteright}s. I.},} {\em J. G{\"o}kova Geom. Topol. GGT} {\bfseries 16} (2023) 1--64, \href{http://arxiv.org/abs/2106.09270}{{ arXiv:2106.09270~[math.AT]}}.

\bibitem{Sugimoto:1999tx}
S.~Sugimoto, {\slshape {Anomaly cancellations in type I D-9 - anti-D-9 system and the USp(32) string theory},} \href{http://dx.doi.org/10.1143/PTP.102.685}{{\em Prog. Theor. Phys.} {\bfseries 102} (1999) 685--699}, \href{http://arxiv.org/abs/hep-th/9905159}{{ arXiv:hep-th/9905159}}.

\bibitem{Bianchi:1991eu}
M.~Bianchi, G.~Pradisi, and A.~Sagnotti, {\slshape {Toroidal compactification and symmetry breaking in open string theories},} \href{http://dx.doi.org/10.1016/0550-3213(92)90129-Y}{{\em Nucl. Phys. B} {\bfseries 376} (1992) 365--386}.

\bibitem{Sen:1997pm}
A.~Sen and S.~Sethi, {\slshape {The Mirror transform of type I vacua in six-dimensions},} \href{http://dx.doi.org/10.1016/S0550-3213(97)81186-8}{{\em Nucl. Phys. B} {\bfseries 499} (1997) 45--54}, \href{http://arxiv.org/abs/hep-th/9703157}{{ arXiv:hep-th/9703157}}.

\bibitem{Witten:1997bs}
E.~Witten, {\slshape {Toroidal compactification without vector structure},} \href{http://dx.doi.org/10.1088/1126-6708/1998/02/006}{{\em JHEP} {\bfseries 02} (1998) 006}, \href{http://arxiv.org/abs/hep-th/9712028}{{ arXiv:hep-th/9712028}}.

\bibitem{Dai:1994kq}
X.-z. Dai and D.~S. Freed, {\slshape {eta invariants and determinant lines},} \href{http://dx.doi.org/10.1063/1.530747}{{\em J. Math. Phys.} {\bfseries 35} (1994) 5155--5194}, \href{http://arxiv.org/abs/hep-th/9405012}{{ arXiv:hep-th/9405012}}. [Erratum: J.Math.Phys. 42, 2343--2344 (2001)].

\bibitem{Zumino:1983rz}
B.~Zumino, Y.-S. Wu, and A.~Zee, {\slshape {Chiral Anomalies, Higher Dimensions, and Differential Geometry},} \href{http://dx.doi.org/10.1016/0550-3213(84)90259-1}{{\em Nucl. Phys. B} {\bfseries 239} (1984) 477--507}.

\bibitem{Alvarez-Gaume:1983ihn}
L.~Alvarez-Gaume and E.~Witten, {\slshape {Gravitational Anomalies},} \href{http://dx.doi.org/10.1016/0550-3213(84)90066-X}{{\em Nucl. Phys. B} {\bfseries 234} (1984) 269}.

\bibitem{Alvarez-Gaume:1984zlq}
L.~Alvarez-Gaume and P.~H. Ginsparg, {\slshape {The Structure of Gauge and Gravitational Anomalies},} \href{http://dx.doi.org/10.1016/0003-4916(85)90087-9}{{\em Annals Phys.} {\bfseries 161} (1985) 423}. [Erratum: Annals Phys. 171, 233 (1986)].

\bibitem{Seiberg:1986by}
N.~Seiberg and E.~Witten, {\slshape {Spin Structures in String Theory},} \href{http://dx.doi.org/10.1016/0550-3213(86)90297-X}{{\em Nucl. Phys. B} {\bfseries 276} (1986) 272}.

\bibitem{Alvarez-Gaume:1986ghj}
L.~Alvarez-Gaume, P.~H. Ginsparg, G.~W. Moore, and C.~Vafa, {\slshape {An O(16) x O(16) Heterotic String},} \href{http://dx.doi.org/10.1016/0370-2693(86)91524-8}{{\em Phys. Lett. B} {\bfseries 171} (1986) 155--162}.

\bibitem{BoyleSmith:2024qgx}
P.~Boyle~Smith and Y.~Zheng, {\slshape {Backfiring bosonisation},} \href{http://dx.doi.org/10.1007/JHEP03(2026)221}{{\em JHEP} {\bfseries 03} (2026) 221}, \href{http://arxiv.org/abs/2403.03953}{{ arXiv:2403.03953~[hep-th]}}.

\bibitem{Heckman:2025wqd}
J.~J. Heckman, J.~McNamara, J.~Parra-Martinez, and E.~Torres, {\slshape {Gliozzi-Scherk-Olive defects: IIA/IIB walls and the surprisingly stable R7-brane},} \href{http://dx.doi.org/10.1103/wqys-zsyf}{{\em Phys. Rev. D} {\bfseries 113} (2026) 066021}, \href{http://arxiv.org/abs/2507.21210}{{ arXiv:2507.21210~[hep-th]}}.

\bibitem{Witten:1982fp}
E.~Witten, {\slshape {An SU(2) Anomaly},} \href{http://dx.doi.org/10.1016/0370-2693(82)90728-6}{{\em Phys. Lett. B} {\bfseries 117} (1982) 324--328}.

\bibitem{Garcia-Etxebarria:2018ajm}
I.~Garc{\'\i}a-Etxebarria and M.~Montero, {\slshape {Dai-Freed anomalies in particle physics},} \href{http://dx.doi.org/10.1007/JHEP08(2019)003}{{\em JHEP} {\bfseries 08} (2019) 003}, \href{http://arxiv.org/abs/1808.00009}{{ arXiv:1808.00009~[hep-th]}}.

\bibitem{Tachikawa:2018njr}
Y.~Tachikawa and K.~Yonekura, {\slshape {Why are fractional charges of orientifolds compatible with Dirac quantization?},} \href{http://dx.doi.org/10.21468/SciPostPhys.7.5.058}{{\em SciPost Phys.} {\bfseries 7} (2019) 058}, \href{http://arxiv.org/abs/1805.02772}{{ arXiv:1805.02772~[hep-th]}}.

\bibitem{Hason:2020yqf}
I.~Hason, Z.~Komargodski, and R.~Thorngren, {\slshape {Anomaly Matching in the Symmetry Broken Phase: Domain Walls, CPT, and the Smith Isomorphism},} \href{http://dx.doi.org/10.21468/SciPostPhys.8.4.062}{{\em SciPost Phys.} {\bfseries 8} (2020) 062}, \href{http://arxiv.org/abs/1910.14039}{{ arXiv:1910.14039~[hep-th]}}.

\bibitem{Vafa:1989ih}
C.~Vafa, {\slshape {Quantum Symmetries of String Vacua},} \href{http://dx.doi.org/10.1142/S0217732389001842}{{\em Mod. Phys. Lett. A} {\bfseries 4} (1989) 1615}.

\bibitem{Gu:2012ib}
Z.-C. Gu and X.-G. Wen, {\slshape {Symmetry-protected topological orders for interacting fermions: Fermionic topological nonlinear {\ensuremath{\sigma}} models and a special group supercohomology theory},} \href{http://dx.doi.org/10.1103/PhysRevB.90.115141}{{\em Phys. Rev. B} {\bfseries 90} (2014) 115141}, \href{http://arxiv.org/abs/1201.2648}{{ arXiv:1201.2648~[cond-mat.str-el]}}.

\bibitem{Bhardwaj:2016clt}
L.~Bhardwaj, D.~Gaiotto, and A.~Kapustin, {\slshape {State sum constructions of spin-TFTs and string net constructions of fermionic phases of matter},} \href{http://dx.doi.org/10.1007/JHEP04(2017)096}{{\em JHEP} {\bfseries 04} (2017) 096}, \href{http://arxiv.org/abs/1605.01640}{{ arXiv:1605.01640~[cond-mat.str-el]}}.

\bibitem{Debray:2026ivi}
A.~Debray, C.~Krulewski, and L.~Stehouwer, {\slshape {Unraveling the Bott spiral},} \href{http://arxiv.org/abs/2605.00316}{{ arXiv:2605.00316~[math-ph]}}.

\bibitem{MR0293642}
J.~Brown, Edgar~H., {\slshape Generalizations of the {K}ervaire invariant,} \href{https://doi.org/10.2307/1970804}{{\em Ann. of Math. (2)} {\bfseries 95} (1972) 368--383}.

\bibitem{Paton:1969je}
J.~E. Paton and H.-M. Chan, {\slshape {Generalized veneziano model with isospin},} \href{http://dx.doi.org/10.1016/0550-3213(69)90038-8}{{\em Nucl. Phys. B} {\bfseries 10} (1969) 516--520}.

\bibitem{Polchinski:1995df}
J.~Polchinski and E.~Witten, {\slshape {Evidence for heterotic - type I string duality},} \href{http://dx.doi.org/10.1016/0550-3213(95)00614-1}{{\em Nucl. Phys. B} {\bfseries 460} (1996) 525--540}, \href{http://arxiv.org/abs/hep-th/9510169}{{ arXiv:hep-th/9510169}}.

\bibitem{Witten:1998cd}
E.~Witten, {\slshape {D-branes and K-theory},} \href{http://dx.doi.org/10.1088/1126-6708/1998/12/019}{{\em JHEP} {\bfseries 12} (1998) 019}, \href{http://arxiv.org/abs/hep-th/9810188}{{ arXiv:hep-th/9810188}}.

\bibitem{Larotonda:2024thv}
V.~Larotonda and L.~Lin, {\slshape {Anomaly inflow and gauge group topology in the 10d Sugimoto string theory},} \href{http://dx.doi.org/10.1007/JHEP06(2025)136}{{\em JHEP} {\bfseries 06} (2025) 136}, \href{http://arxiv.org/abs/2412.17894}{{ arXiv:2412.17894~[hep-th]}}.

\bibitem{Kapustin:1999di}
A.~Kapustin, {\slshape {D-branes in a topologically nontrivial B field},} \href{http://dx.doi.org/10.4310/ATMP.2000.v4.n1.a3}{{\em Adv. Theor. Math. Phys.} {\bfseries 4} (2000) 127--154}, \href{http://arxiv.org/abs/hep-th/9909089}{{ arXiv:hep-th/9909089}}.

\bibitem{Gao:2010ava}
D.~Gao and K.~Hori, {\slshape {On The Structure Of The Chan-Paton Factors For D-Branes In Type II Orientifolds},} \href{http://arxiv.org/abs/1004.3972}{{ arXiv:1004.3972~[hep-th]}}.

\bibitem{BoyleSmith:2026oay}
P.~Boyle~Smith and Y.~Tachikawa, {\slshape {On Generalised Discrete Torsion},} \href{http://arxiv.org/abs/2604.01225}{{ arXiv:2604.01225~[hep-th]}}.

\bibitem{Beaudry:2018ifm}
A.~Beaudry and J.~A. Campbell, {\slshape {A Guide for Computing Stable Homotopy Groups},} \href{http://arxiv.org/abs/1801.07530}{{ arXiv:1801.07530~[math.AT]}}.

\bibitem{Wan:2018bns}
Z.~Wan and J.~Wang, {\slshape {Higher anomalies, higher symmetries, and cobordisms I: classification of higher-symmetry-protected topological states and their boundary fermionic/bosonic anomalies via a generalized cobordism theory},} \href{http://dx.doi.org/10.4310/AMSA.2019.v4.n2.a2}{{\em Ann. Math. Sci. Appl.} {\bfseries 4} (2019) 107--311}, \href{http://arxiv.org/abs/1812.11967}{{ arXiv:1812.11967~[hep-th]}}.

\bibitem{Wan:2019soo}
Z.~Wan, J.~Wang, and Y.~Zheng, {\slshape {Higher anomalies, higher symmetries, and cobordisms II: Lorentz symmetry extension and enriched bosonic / fermionic quantum gauge theory},} \href{http://dx.doi.org/10.4310/AMSA.2020.v5.n2.a2}{{\em Ann. Math. Sci. Appl.} {\bfseries 05} (2020) 171--257}, \href{http://arxiv.org/abs/1912.13504}{{ arXiv:1912.13504~[hep-th]}}.

\bibitem{Adams1958}
J.~F. Adams, {\slshape On the structure and applications of the steenrod algebra,} \href{https://doi.org/10.1007/BF02564578}{{\em Commentarii Mathematici Helvetici} {\bfseries 32} (1958) 180--214}.

\bibitem{ABP67}
D.~W. Anderson, E.~H. Brown, and F.~P. Peterson, {\slshape The structure of the spin cobordism ring,} \href{http://www.jstor.org/stable/1970690}{{\em Annals of Mathematics} {\bfseries 86} (1967) 271--298}.

\end{thebibliography}\endgroup
\end{document}